\documentclass[aps,twocolumn,floatfix,altaffilletter,superscriptaddress,preprintnumbers,
               tightenlines,showpacs,showkeys,notitlepage,nofootinbib,
               longbibliography]{revtex4-2}

\usepackage{amsmath,amssymb} % Maths
\usepackage{graphicx} % Figures
\usepackage{tabularx}
\usepackage{fancybox}
\usepackage[export]{adjustbox} % Vertical alignment of figures
\usepackage{url} % URL formatting
\usepackage[dvipsnames]{xcolor} % Colors
\usepackage{siunitx} % Proper unit formating
\usepackage{orcidlink} % for ORCIDs behind author names
\usepackage{tabularx,booktabs} % Tables
\usepackage{titlesec}
\usepackage{microtype}
\usepackage{dcolumn}% Align table columns on decimal point
\usepackage{bm,bbm}% bold math

\allowdisplaybreaks

\makeatletter
\def\thesection{\arabic{section}}%
\def\p@section{}%
\def\thesubsection{\thesection.\arabic{subsection}}%
\def\p@subsection{}%
\def\thesubsubsection{\thesubsection.\arabic{subsubsection}}%
\def\p@subsubsection{}%
\def\appendix{%
    \par
    \setcounter{section}\z@
    \setcounter{subsection}\z@
    \setcounter{subsubsection}\z@
    \def\thesubsection{\thesection.\arabic{subsection}}%
    \def\thesubsubsection{\thesubsection.\arabic{subsubsection}}%
    \def\p@subsection{}%
    \def\p@subsubsection{}%
    \@addtoreset{equation}{section}%
    \def\theequation@prefix{\thesection}%
    \addtocontents{toc}{\protect\appendix}%
    \@ifstar{%
    \def\thesection{\unskip}%
    \def\theequation@prefix{A.}%
    }{%
    \def\thesection{\Alph{section}}%
    }%
}%
\makeatother

\titleformat*{\section}{\centering\bfseries\MakeUppercase}
\titlelabel{\thetitle\quad}
\titleformat*{\paragraph}{\bfseries}
\titlespacing*{\paragraph}{0pt}{3.25ex plus 1ex minus .2ex}{1em}
\titlelabel{\thetitle\quad}  % remove dot behind section number

\usepackage{hyperref} % References
\usepackage[capitalize]{cleveref} % Nice referencing
\newcommand\myshade{80} % Reference coloring
\colorlet{mylinkcolor}{ForestGreen}
\colorlet{mycitecolor}{Red}
\colorlet{myurlcolor}{violet}
\hypersetup{
  linkcolor  = mylinkcolor!\myshade!black,
  citecolor  = mycitecolor!\myshade!black,
  urlcolor   = myurlcolor!\myshade!black,
  colorlinks = true
}

\definecolor{QCcolor}{HTML}{ffffee}

\renewcommand{\vec}[1]{{\mathbf{#1}}}
\newcommand{\bra}[1]{\ensuremath{\langle #1 |}}   % Bra vector
\newcommand{\ket}[1]{\ensuremath{| #1 \rangle}}   % Ket vector
\newcommand{\amp}[3]{\ensuremath{\left\langle #1 \,\left|\, #2%
                     \,\right|\, #3 \right\rangle}}  % QM amplitude
\begin{document}

\title{Quantum Simulation of Collective Neutrino Oscillations using Dicke States}

\newcommand{\utoronto}{Department of Physics, University of Toronto, Toronto, ON M5S 1A7, Canada}
\newcommand{\cern}{CERN, 1211 Geneva 23, Switzerland}
\newcommand{\jgu}{PRISMA Cluster of Excellence \& Mainz Institute for
                  Theoretical Physics, \\
                  Johannes Gutenberg University, 55099 Mainz, Germany}
\newcommand{\tifr}{Tata Institute of Fundamental Research, Homi Bhabha Road, Colaba, Mumbai 400005, India}

\author{Katarina Bleau \orcidlink{0000-0003-0600-4996}\,}
\email{kbleau@uni-mainz.de}
\affiliation{\jgu}

\author{Nikolina Ilic \orcidlink{0000-0003-0105-7634}\,}
\email{nikolina.ilic@cern.ch}
\affiliation{\utoronto}

\author{Joachim Kopp \orcidlink{0000-0003-0600-4996}\,}
\email{jkopp@cern.ch}
\affiliation{\jgu}

\author{Ushak Rahaman \orcidlink{0000-0003-0419-2970}\,}
\email{ushak.rahaman@cern.ch}
\affiliation{\utoronto}
\affiliation{\tifr}

\author{Xin Yue Yu \orcidlink{0000-0002-6976-2703}\,}
\email{xyz.yu@mail.utoronto.ca}
\affiliation{\utoronto}

\date{\today}
\preprint{MITP-YYY} 
\keywords{}

%=============================================================================

\begin{abstract}
\noindent
In dense neutrino gases, which exist for instance in supernovae, the flavour states of different neutrinos may become entangled with one another. The theoretical description of such systems may therefore call for simulations on a quantum computer. Existing quantum simulations of simple toy systems are not optimal in the sense that they do not fully exploit the symmetries of the system. Here, we propose a new class of qubit-efficient algorithms based on Dicke states and the $su(2)$ spin algebra. We demonstrate the excellent performance of these algorithms both on classical and on quantum hardware.
\end{abstract}

%=============================================================================
\maketitle
%=============================================================================

\textbf{Introduction.} Neutrinos can justifiably be called the shape-shifters of the elementary particle zoo, given the intricate ways in which the three neutrino types (or flavours) change into one another during propagation. This phenomenon, called neutrino oscillations, has been well established for neutrinos from the Sun \cite{art:SolarNeutrino}, the upper atmosphere \cite{art:AtmoNuSuper-Kamiokande:1998kpq}, and human-made neutrino sources \cite{KamLAND:2002uet, MINOS:2011amj}. It is moreover well established that neutrino oscillation patterns change in the presence of ambient matter -- a phenomenon known as the Mikheyev--Smirnov--Wolfenstein (MSW) effect \cite{Mikheyev:1985zog, Wolfenstein:1977ue}. At the core of this effect is \emph{coherent forward scattering}, wherein the neutrino flavour may change, but no momentum transfer occurs. The situation becomes yet more intriguing when the ambient matter consists of neutrinos, too. This is the case for instance deep inside core-collapse supernova explosions, binary neutron star mergers, or the early Universe \cite{Pantaleone:1992eq, Pantaleone:1992xh, Qian:1994wh, Samuel:1995ri, Pastor:2002we, Balantekin:2004ug}. In all of these systems, the neutrino density is so high that even the feeble neutrino--neutrino interactions become non-negligible.

A number of authors have  argued that neutrino--neutrino interactions in dense neutrino gases may be further complicated by quantum entanglement. In fact, when neutrinos interact with one another, their flavours become entangled, and it stands to reason that such flavour entanglement might have phenomenological consequences. Others have, however, argued that this is \emph{not} the case \cite{Friedland:2003dv, Friedland:2003eh, Shalgar:2023ooi, Johns:2023ewj} at the very high neutrino density, $n_\nu \sim \SI{e30}{cm^{-3}}$, realized in a supernova. Yet it is known that the opposite conclusion holds for more rarefied systems \cite{Cervia:2019res, Shalgar:2023ooi, Johns:2023ewj}. Notably, in more dilute systems, the Boltzmann approach \cite{Sigl:1993ctk} may break down, and it may become necessary to consider full multiparticle evolution. This would entail a computational complexity that scales exponentially with the number of simulated neutrinos, $N$. Such systems are therefore a prime application for \emph{quantum computing}, true to the reasoning that a highly entangled quantum system is best simulated on a device that is itself a highly entangled quantum system. Pioneering work on this topic has been carried out in refs.~\cite{
    Hall:2021rbv,       % Roggero et al., collective nu oscillations on a quantum computer
    Yeter-Aydeniz:2021olz, % collective nu osc on a quantum computer
    Jha:2021itm,        % Hyderabad group
    Amitrano:2022yyn,   % Roggero et al., Trapped-ion quantum simulation
    Balantekin:2023qvm, % Balantekin et al., QIS / quantum simulation in neutrino physics
    Siwach:2023wzy,     % Balantekin et al., hybrid quantum-classical algorithm
    Turro:2024shh,      % qubits and qutrits
    Chernyshev:2024pqy, % Savage et al., quantum magic in the neutrino sector
    Singh:2024vpu,      % IIT Mohali group, simulations on NMR quantum processor
    Spagnoli:2025etu,   % Roggero et al., 3-flavors, qutrits
    Tripathi:2025cok}.  % Mumbai group

Here, we will not enter the discussion on the relevance of quantum entanglement in realistic environments (we defer this to upcoming work \cite{Bleau:inprep}), but we will focus on the development of novel algorithms for the simulation of toy systems in which entanglement is known to be important. Exploiting the symmetries of these systems, our algorithms will use significantly fewer qubits than other approaches, albeit in some cases at the expense of increased circuit depth.

%=============================================================================

\begin{figure}
    \centering
    \includegraphics[width=1.\columnwidth]{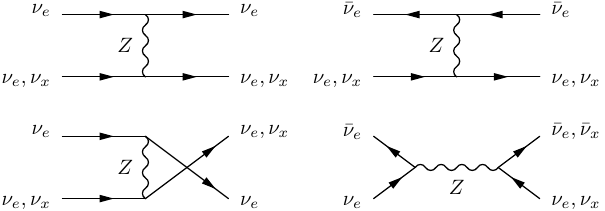}
    \caption{Representative microscopic processes corresponding to the neutrino--neutrino interactions (left) and neutrino--antineutrino interactions (right).}
    \label{fig:feynman-diag}
\end{figure}

\textbf{Collective Neutrino Oscillations.}
Neutrino--neutrino coherent forward scattering in ultradense environments is governed by the interactions depicted in \cref{fig:feynman-diag}. We work here in the two-flavour approximation, subsuming the $\nu_\mu$ and $\nu_\tau$ flavours into $\nu_x$. This simplification is motivated by the fact that $\nu_\mu$ and $\nu_\tau$ behave similarly in supernovae. The two-neutrino and neutrino--antineutrino interaction Hamiltonians corresponding to the diagrams in \cref{fig:feynman-diag} read
\begin{align}
    H_{\nu\nu} &= \sqrt{2} G_F n_\nu (1 - \cos\alpha)
                  \begin{pmatrix}
                      2 \\
                        & 1 & 1 \\
                        & 1 & 1 \\
                        &   &   & 2
                  \end{pmatrix}
                  &{\color{gray}\!\!\!\!
                  \begin{array}{l}
                      \ket{\nu_e\nu_e} \\
                      \ket{\nu_e\nu_x} \\
                      \ket{\nu_x\nu_e} \\
                      \ket{\nu_x\nu_x}
                  \end{array}}
    \label{eq:H-nunu} \\
    H_{\bar\nu\nu} &= \sqrt{2} G_F n_\nu (1 - \cos\alpha)
                      \begin{pmatrix}
                          -2 &        &        & \!\!\!-1\\
                             & \!\!\!-1 \\
                             &        & \!\!\!-1 \\
                          -1 &        &        & \!\!\!-2
                      \end{pmatrix}
                      &{\color{gray}\!\!\!\!
                      \begin{array}{l}
                          \ket{\bar\nu_e\nu_e} \\
                          \ket{\bar\nu_e\nu_x} \\
                          \ket{\bar\nu_x\nu_e} \\
                          \ket{\bar\nu_x\nu_x}
                      \end{array}}
    \label{eq:H-nubarnu}
\end{align}
where $G_F$ is Fermi's constant, $n_\nu$ is the neutrino number density, and $\alpha$ is the angle between the momentum vectors of the two neutrino modes. In this letter, we limit ourselves to the single angle approximation, where $\alpha$ remains the same for all interacting (anti-) neutrinos. In grey, we indicate the basis in which these matrices are written. Each neutrino also experiences standard vacuum oscillations, governed by the Hamiltonian
\begin{align}
    H_0 = U \begin{pmatrix}
                0 \\
                  & \frac{\delta m^2}{2 E}
            \end{pmatrix} U^\dag ,
    \label{eq:H0}
\end{align}
where $\delta m^2 \equiv m_2^2 - m_1^2$ is the mass squared difference, $E$ is the neutrino energy, and $U$ is the $2 \times 2$ mixing matrix parameterized by a mixing angle $\theta$. The effect of ordinary background matter can be taken into account by shifting $\theta$ and $\delta m^2$ to their effective in-medium values \cite{Akhmedov:1999uz}.

%=============================================================================

\textbf{Conventional quantum circuits for collective neutrino oscillations.} In the literature on quantum simulations of collective neutrino oscillations, an ensemble of $N$ neutrinos in the two-flavour approximation is represented by a system of $N$ qubits. The flavour eigenstates $\ket{\nu_e}$ and $\ket{\nu_x}$ of each neutrino are mapped onto the states of the corresponding qubit according to $\ket{\nu_e} \leftrightarrow \ket{0}$, $\ket{\nu_x} \leftrightarrow \ket{1}$. (The formalism can be extended to three flavours by either using two qubits per neutrino, or by using qutrits \cite{Turro:2024shh, Spagnoli:2025etu}.) The system is evolved over time according to the Hamiltonian
\begin{align}
    H_q \equiv H_{q,0} + H_{q,\nu\nu}
    \label{eq:H-q}
\end{align}
with
\begin{align}
    H_{q,0} \equiv \frac{1}{2} \sum_p \vec{b}_p \cdot \hat{\vec\sigma}_p
    \quad\text{and}\quad
    H_{q,\nu\nu} \equiv \sum_{p<q} J_{pq} \hat{\vec\sigma}_p \otimes \hat{\vec\sigma}_q .
    \label{eq:Hq0-Hqnunu}
\end{align}
The first term, which is equivalent to \cref{eq:H0}, describes vacuum oscillations and MSW interactions with ordinary matter. It contains the vector of Pauli matrices acting on the $p$-th qubit, $\hat{\vec\sigma}_p \equiv (\hat\sigma_p^x, \hat\sigma_p^y, \hat\sigma_p^z)$, as well as $\vec{b}_p \equiv \Delta_p (\sin 2\theta, 0, -\cos 2\theta)$, where $\Delta_p = \delta m^2 / (2 E_p)$ is the oscillation frequency of the $p$-th neutrino and $\theta$ is the mixing angle. The second term in \cref{eq:Hq0-Hqnunu} is equivalent (up to terms proportional to the identity matrix) to \cref{eq:H-nunu}. It is parameterized by the self-interaction strength $J_{pq} \equiv (G_F n_\nu / \sqrt{2}) (1 - \cos\alpha_{pq})$, with $\alpha_{pq}$ the angle between the momentum vectors of modes $p$ and $q$. For neutrino--antineutrino interactions, the self-interaction term is replaced by
\begin{align}
    H_{q,\nu\bar\nu} \equiv
        \sum_{p<q} J_{pq} \big( - \hat\sigma_p^x \otimes \hat\sigma_q^x
                                + \hat\sigma_p^y \otimes \hat\sigma_q^y
                                - \hat\sigma_p^z \otimes \hat\sigma_q^z \big)
    \label{eq:H-q-nubarnu}
\end{align}
Note that $H_{q,\nu\bar\nu} = (\mathbbm{1}_2 \otimes \hat\sigma_2) \cdot H_{q,\nu\nu} \cdot (\mathbbm{1}_2 \otimes \hat\sigma_2)^\dag$. Therefore, $\nu\nu$, $\nu\bar\nu$, and $\bar\nu\bar\nu$ interactions can all be described by $H_{q,\nu\nu}$ alone if we change the mapping between flavour eigenstates and qubit states for anti-neutrinos to $\ket{\bar\nu_e} \leftrightarrow \ket{1}$, $\ket{\bar\nu_x} \leftrightarrow -\ket{0}$, and if we change $\vec{b}_p \to -\vec{b}_p$ for anti-neutrinos.

As exponentiating the full $2^N \times 2^N$ matrix $H_q$ is unfeasible for large $N$, the $N$-qubit quantum state $\ket{\Psi} \equiv \ket{\psi_i} \otimes \ldots \otimes \ket{\psi_n}$ is evolved over time using a second-order Suzuki--Trotter decomposition:
\begin{align}
    \ket{\Psi(t)} &= \prod_{j=1}^{n_\text{steps}} \prod_{p<q} e^{-i h_{pq} t/n_\text{steps}} \, \ket{\Psi(0)}
    \label{eq:Trotter}
\intertext{with}
    h_{pq} &\equiv \frac{1}{2} \frac{\vec{b}_p \cdot \hat{\vec\sigma}_p + \vec{b}_q \cdot \hat{\vec\sigma}_q}{n_\text{steps}-1}
                 + J_{pq} \hat{\vec\sigma}_p \otimes \hat{\vec\sigma}_q.
    \label{eq:hpq}
\end{align}
Here, $n_\text{steps}$ is the number of time steps. In practice, \cref{eq:Trotter} can be implemented on a universal quantum computer using $R_X$ and $R_Z$ gates for the standard oscillation term, and $R_{XX}$, $R_{YY}$, and $R_{ZZ}$ gates for the self-interaction term. The latter can often be further optimized for the specific set of basis gates available on a given quantum processing unit (QPU). Details on the circuit implementation and optimization are given in \cref{app:conventional}.

%=============================================================================

\textbf{Dicke States.}
If the coupling strengths $\vec{b}_p$ and $J_{pq}$ appearing in the Hamiltonian are not all distinct, specifically when neutrinos can be grouped into ensembles of particles that interact in the same way, the system exhibits an extended degree of symmetry that can be exploited to optimize the simulation. In particular, the Hamiltonian in \cref{eq:H-q} is invariant under permutations of neutrinos that share the same 4-momentum, i.e.\ the same $\vec{b}_p \equiv \vec{b}$ and $J_{pq} \equiv J$. To exploit this symmetry it is useful to remind ourselves that a system of $N$ qubits is effectively a system of $N$ spin-$\frac{1}{2}$ particles. The $su(2)$ algebra is then a powerful tool for understanding the system. Given that $\frac{1}{2} \hat{\vec\sigma}_p$ is the spin operator corresponding to the $p$-th qubit, the qubit Hamiltonian from \cref{eq:H-q,eq:Hq0-Hqnunu} for $\vec{b}_p = \vec{b}$, $J_{pq} = J$ can be rewritten as
\begin{align}
    H_s \equiv \vec{b} \cdot \vec{S}_\text{tot}
             + 2 \, J \, \vec{S}_\text{tot}^2,
    \label{eq:Hs-1-mode}
\end{align}
with $\vec{S}_\text{tot}$ the total spin vector. We have dropped a term $-N J \vec{s}^2$, where $\vec{s}$ is the spin vector of a single qubit ($\vec{s}^2 = \frac{1}{2} (\frac{1}{2} + 1)$), as it is proportional to the identity matrix and will hence only contribute an overall phase. Note that $\vec{S}_\text{tot}^2$ is a constant of motion. Therefore, if the system is initially in an eigenstate of $\vec{S}^{2}_\text{tot}$ and $S_\text{tot}^z$, a so-called Dicke state, the second term can also be dropped. The vacuum oscillation term, $\vec{b} \cdot \vec{S}_\text{tot}$, describes a spin precessing around the vector $\vec{b}$. Clearly, time evolution under \cref{eq:Hs-1-mode} can be very easily calculated on classical hardware.

For multiple neutrino modes with different energies and propagation directions, and/or with mixtures of neutrinos and anti-neutrinos (the latter again implemented as described below \cref{eq:H-q-nubarnu}), \cref{eq:Hs-1-mode} generalizes to
\begin{align}
    H_s \equiv \sum_i \vec{b}_i \cdot \vec{S}_i
             + 2 \sum_i J_i \vec{S}_i^2 
             + 4 \sum_{i<j} J_{ij} \vec{S}_i \cdot \vec{S}_j,
    \label{eq:Hs-n-modes}
\end{align}
where $\vec{b}_i$ parameterizes vacuum oscillations (+ MSW matter effects with ordinary matter) of the $i$-th mode, $J_i$ is the self-interaction strength among neutrinos in mode $i$, and $J_{ij}$ is the interaction strength between modes $i$ and $j$. We have again dropped a term $N \vec{s}^2$ which is proportional to the identity matrix. For the same reason we can also drop the $J_i$ term if each mode starts out in an eigenstate of the corresponding $\vec{S}_i^2$.

Under the same condition, we see that the complexity of the Hilbert space on which \cref{eq:Hs-n-modes} acts grows exponentially with the number of \emph{modes} (and only linearly with the number of neutrinos per mode), while the complexity of the original Hamiltonian, \cref{eq:H-q}, is exponential in the number of neutrinos. This presents a significant computational advantage, which we can exploit both on classical and on quantum hardware. It also implies that whether or not quantum advantage is realized in the Dicke approach depends on the number of neutrino modes, not the total number of neutrinos.

%=============================================================================

\textbf{Encoding Dicke states in quantum registers.}
We now describe how the Dicke state Hamiltonian from \cref{eq:Hs-n-modes} can be implemented on a quantum computer in a qubit-efficient way. We will in the following always assume that for a mode comprised of $N_i$ neutrinos, the initial state is an eigenstate of the corresponding $\vec{S}_i^2$ with quantum numbers $S_i = N_i/2$ and $m_i = \pm S_i$ (i.e.\ all neutrinos in the mode are initially in the same flavor). If this is not the case, the mode can be split into multiple new modes, each of which satisfies the condition. Permutation symmetry then restricts the Hilbert space of the $i$-th mode to a subspace spanned by $N_i+1$ states $\ket{S_i, m_i}$ with $m_i = -S_i, -S_i+1, \ldots, +S_i$. Such a state can be encoded as a binary number in a register consisting of $\lceil \log_2(N_i+1) \rceil$ qubits via the computational-basis mapping
\begin{align}
    \ket{j}_{\text{comp}} \;\longleftrightarrow\;
    \ket{S,\; m = j - S},
    \label{eq:log_encoding}
\end{align}
with $j \in \{0, 1, \ldots, N\}$. Any such state can be straightforwardly prepared from a state in which all qubits are in the $\ket{0}$ state by applying a series of Pauli $X$ gates. For multimode systems, each mode is encoded in its own quantum register.

\begin{figure*}
    \centering
    \begin{tabular}{c@{\quad\quad}c}
         \begin{minipage}{0.68\textwidth}
             \colorbox{QCcolor}{\includegraphics[width=\textwidth]{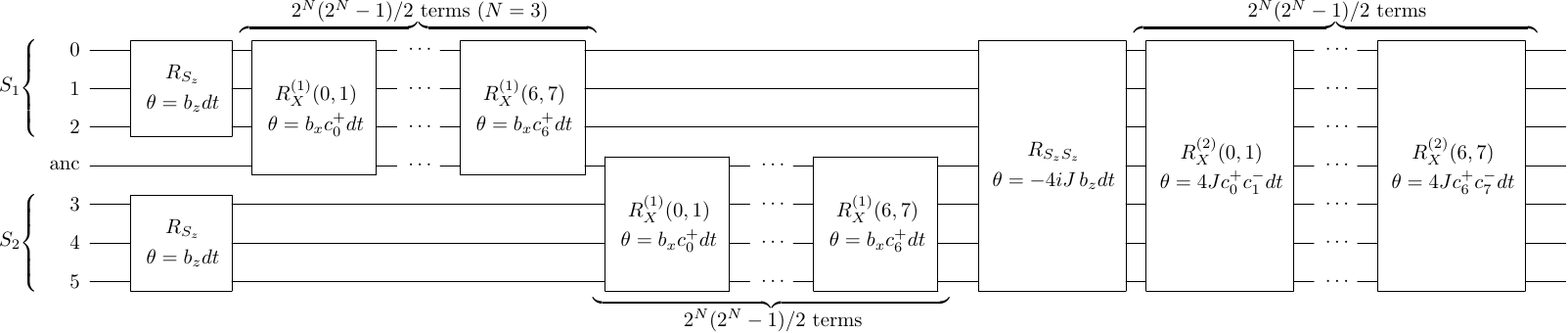}}
         \end{minipage}
         &
         \begin{minipage}{0.28\textwidth}
             \includegraphics[width=\textwidth]{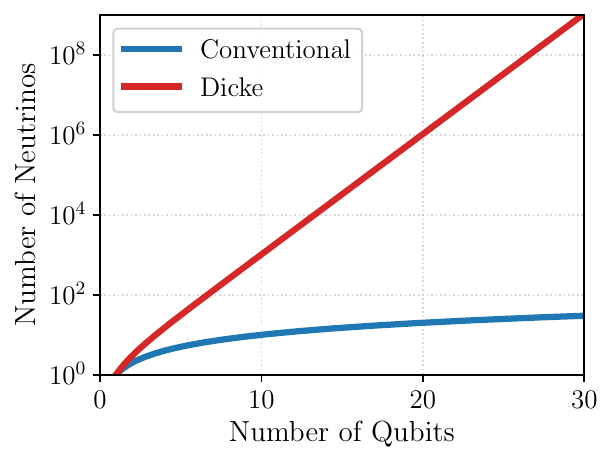}
         \end{minipage}
         \\
         (a) & (b)
    \end{tabular}
    \caption{(a) Quantum circuit for a single Trotter step of a system of two interacting neutrino modes $S_1$ and $S_2$ encoded as qubit-efficient Dicke states with three qubits each. The first part of the circuit, consisting of operations on a single register (+ one ancillary qubit), implements the standard oscillation term in the Hamiltonian, and the second part, composed of gates spanning both registers (+ the ancillary) represent the self-interaction term.
    (b) Size of the physical system represented as a function of the number of qubits required for different implementations.}
    \label{fig:trotter-step-circuit}
\end{figure*}

As shown in \cref{fig:trotter-step-circuit}~(a), implementing the Trotterized time evolution of the system, governed by the Hamiltonian in \cref{eq:Hs-n-modes}, requires generalizations of the $R_X$, $R_Z$, $R_{XX}$, $R_{YY}$, and $R_{ZZ}$ gates to Dicke states. Here, we explain these briefly; full details are given in \cref{app:full-dicke}. The implementation is most straightforward for the operation $R_{S_z} \equiv \exp(-i b_{i,z} dt \, \hat{S}_{i,z})$, which corresponds to one $R_Z$ gate for each qubit in the $i$-th register, where the rotation angle for the $p$-th qubit is weighted with a factor $2^p$.

The operation $\exp(-i b_{i,x} dt \, \hat{S}_{i,x})$ is more involved. We use the identity
\begin{multline}
    \exp\!\big(\!-i b_{i,x} \,dt\, \hat{S}_{i,x} \big)
        = \exp\!\bigg(\!\!-i \frac{b_{i,x}}{2} \,dt\,(\hat{S}_i^+ + \hat{S}_i^-)\!\bigg) \\
        = \!\!\prod_{k=0}^{N_i-1}\!\! \exp\!\bigg(\!\!\!-i \frac{b_{i,x}}{2} c_k \,dt\, \hat{X}_{k,k+1} \!\bigg)
        \equiv \!\!\prod_{k=0}^{N_i-1} R_X^{(1)}(k,k+1) ,
    \label{eq:RX1-gate}
\end{multline}
where $\hat{S}_i^\pm = \hat{S}_{i,x} \pm i \hat{S}_{i,y}$ are the ladder operators, $\hat{X}_{k,k+1}$ is a Pauli $X$ operation acting between the $\ket{k}$ and $\ket{k+1}$ states of the quantum register, and $c_k$ is a prefactor determined by the $su(2)$ algebra. \Cref{eq:RX1-gate} can be implemented in two ways (see \cref{app:full-dicke}, \cref{fig:rx-gate-circuit,fig:rx-gate-circuit-alt} for details): either by entangling the register's ($\ket{k}$, $\ket{k+1}$) subspace onto the $\ket{0}$ and $\ket{1}$ states of an ancillary qubit and carrying out an ordinary $R_X$ rotation on the ancilla; or by writing the ladder operations $\ket{k+1} \bra{k}$, $\ket{k} \bra{k+1}$ as sums of Pauli strings (Kronecker products of the single-qubit operations $\hat{I}$, $\hat{X}$, $\hat{Y}$, $\hat{Z}$), whose exponentials can be evaluated using a general algorithm. 

For the operation $R_{S_z S_z} \equiv \exp\big[ -i \sum_{j<k} J_{jk} \hat{S}_{j,z} \otimes \hat{S}_{k,z} \big]$, the main ingredient is a series of controlled $CR_Z$ gates between all qubit pairs, with the rotation angle of the gate acting on the $p$-th qubit of the first register, and the $q$-th qubit of the second register scaled by $2^{p+q}$.

Finally, for $\exp\big[ -i \sum_{j<k} J_{jk} \big( \hat{S}_{j,x} \otimes \hat{S}_{k,x} + \hat{S}_{j,y} \otimes \hat{S}_{k,y} \big)\big]$, we follow similar ideas as in \cref{eq:RX1-gate} above: we first write $S_{j,x}$, $S_{j,y}$ in terms of ladder operators. In each two-dimensional subspace of the two-register system, the operation then reduces to $R_X$. We carry out this generalized $R_X$ (which we dub $R_X^{(2)}$) either with the help of an appropriately conditioned ancillary qubit, or using Pauli strings. Full details are again given in \cref{app:full-dicke}.

\begin{figure}
    \centering
    \includegraphics[width=1\linewidth]{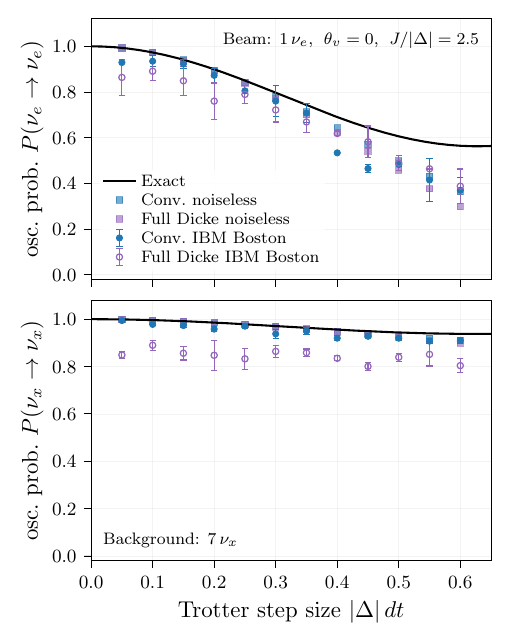}
    \caption{Time-evolution of a system of one ``beam'' $\nu_e$ (top) interacting with seven ``background'' $\nu_x$ (bottom) over a single Trotter step. We compare the conventional implementation using one qubit per neutrino (8 qubits total) to our qubit-efficient method based on Dicke state encoding (4 qubits + 1 ancilla).}
    \label{fig:single-Trotter}
\end{figure}

The performance of the algorithm is illustrated in \cref{fig:single-Trotter}, where we show the results of a single Trotter step for a system of one ``beam'' $\nu_e$ interacting with seven ``background'' $\nu_x$.  The simulations were run on the IBM Boston device with a Heron~r3 QPU, using 1\,024 shots and default error mitigation options (see \cref{app:error_mitigation} for details). We see that the conventional encoding (8 qubits) and our Dicke state encoding (4 qubits + 1 ancilla) perform similarly. The $\mathcal{O}(10\%)$ error in the survival probability of the background neutrinos for the Dicke method arises because the Dicke encoding is disproportionately sensitive to flips of the higher-significance qubits in each register. This, combined with the larger gate count highlights the fact that the Dicke method is optimal in a low-noise environment with a limited number of qubits (for instance trapped ion systems \cite{Haeffner:2005bpb, Haeffner:2008jjg, Monroe:2019asq}, neutral atoms \cite{Browaeys:2020kzz, Saffman:2016kig}, protected bosonic modes in superconducting cavities \cite{Mirrahimi:2013sgk}, spin-based solid-state qubits \cite{Muhonen:2014gsv, Veldhorst:2014tsi}, or hypothetical topological quantum computers \cite{Kitaev:1997wr, Nayak:2008}), while the conventional encoding has advantages on large but noisy devices.

%=============================================================================

\textbf{Diagonal subspace reduction for bipolar systems.}
We now consider a symmetric bipolar system with an initial state consisting of $N$ electron neutrinos and $N$ electron antineutrinos, i.e.\ initially $m_1 = -m_2 = -N/2$. This system has an enhanced degree of symmetry compared to the general case, allowing for yet more compression. The Hamiltonian now simplifies to
\begin{align}
    H_s \equiv \sum_{i=1,2} \vec{b}_i \cdot \vec{S}_i
             + 4 J \vec{S}_1 \cdot \vec{S}_2 \,.
    \label{eq:H-Dicke-bipolar}
\end{align}
Using again the $su(2)$ ladder operators $S_i^\pm = (S_{i,x} \pm i S_{i,y})$, which act as
\begin{align}
    S^\pm \ket{S,m} &= \sqrt{(S(S+1) - m(m\pm1)} \, \ket{S,m\pm1},
\end{align}
the self-interaction term can be rewritten as
\begin{align}
    4 J_{12} \big(
        S_{1,z} S_{2,z} + \tfrac{1}{2} S_1^+ S_2^- + \tfrac{1}{2} S_1^- S_2^+
    \big) \,.
\end{align}
This term preserves the condition $m_1 = -m_2$. The standard oscillation term violates it, but standard oscillations are often negligible, notably when the mixing angle is small such that $b_{i,x} \simeq 0$, or when $|J_{12}| \gg |\Delta = \delta m^2 / (2 E)|$. If we drop the $b_{i,x}$ term from $H_s$, the Hamiltonian operates only on the antidiagonal space spanned by joint Dicke states of the form
\begin{align}
    \big\{ \ket{m,-m} :
        m = -S, \ldots, +S\}
        \quad\text{with}\quad S = N/2.
    \label{eq:diag_subspace}
\end{align}
Physically, this follows from the fact that the $s$-channel scattering responsible for neutrino--antineutrino flavour changes $(\nu_e \bar{\nu}_e \leftrightarrow \nu_x \bar{\nu}_x)$ always produces a $\nu$--$\bar\nu$ pair of the same flavour. The resulting Hamiltonian in this subspace is (anti-)tridiagonal:
\begin{align}
    \amp{m,-m}{H_s}{m,-m}
        &= - 2\Delta\cos 2\theta\; m - 4 J\, m^2,
        \label{eq:H_diag} \\
    \amp{m\!-\!1,-m\!+\!1}{H_s}{m,-m}
        &= 2 J \,(S+m)(S-m+1) \,.
        \label{eq:H_offdiag}
\end{align}
This reduces the Hilbert space dimension from $(N+1)^2$ to $N+1$.

On a quantum computer, we can therefore encode the bipolar system using a single register. In analogy to the discussion above, the diagonal part of the Hamiltonian, \cref{eq:H_diag}, can be implemented using a combination of $R_Z$ and $R_{ZZ}$ gates, and the off-diagonal part, \cref{eq:H_offdiag}, can be written in terms of ladder operators. See \cref{app:diagonal-dicke} for details.

\begin{figure}
    \centering
    \includegraphics[width=1\linewidth]{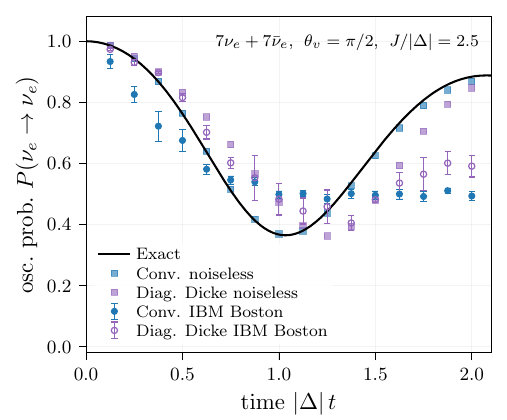}
    \caption{Time-evolution over 16 Trotter steps of a bipolar system of $7 \nu_e + 7 \bar\nu_e$, all of the same energy. We compare the conventional implementation (14 qubits) to the ``diagonal Dicke'' method introduced here (3 qubits). We see that the conventional method becomes noise dominated sooner, while the diagonal Dicke method suffers from a somewhat larger Trotter error.}
    \label{fig:comparison}
\end{figure}

Results are shown in \cref{fig:comparison}, where we follow the time evolution of a bipolar system with $7 \nu_e + 7 \bar\nu_e$ over 16 Trotter steps. The diagonal Dicke encoding not only requires fewer qubits (3) than the conventional encoding (14), but it also becomes noise-dominated significantly later: the open purple data points trace out the oscillation dip, while the filled blue data points level out at 0.5 after about seven Trotter steps. The diagonal Dicke method exhibits a larger Trotterization error (compare the solid purple and blue squares) since it employs a first-order Trotter approximation, compared to the second-order Trotterization in the conventional method. Note that we have chosen a vanishing vacuum mixing angle here to avoid further errors in the (approximate) vacuum oscillation term.

%=============================================================================

\textbf{Outlook.} Having demonstrated how the symmetries of collectively oscillating neutrino systems can be exploited to render quantum simulations of such systems more efficient, a possible next step is an extension to three neutrino flavours using qutrit instead of qubit registers and exploiting the properties of the $su(3)$ rather than the $su(2)$ algebra \cite{Turro:2024shh, Spagnoli:2025etu}. As an alternative to hardware qutrits, it is also possible to create logical qutrits (composed of two qubits each), but at the expense of further adding to the circuit complexity. An interesting alternative could be qudits or qumodes (quantum systems with $n \gg 2$ possible states) \cite{Heeres:2015cnj, Stavenger:2022wzz, Araz:2024dcy}, in which case a full neutrino mode could be represented by a single qudit.

%=============================================================================

\textbf{Acknowledgments.} It is a pleasure to thank Basudeb Dasgupta and Enrique Rico Ortega for very insightful discussions. We are moreover grateful to CERN, where crucial parts of this work were carried out, for hospitality and for access to IBM QPUs through the \href{quantum.cern}{CERN Quantum Technology Initiative}. KB and JK have received support from the Cluster of Excellence \href{https://www.prisma.uni-mainz.de/}{``Precision Physics, Fundamental Interactions and Structure of Matter``}, PRISMA++ (EXC 2118/2, Project ID 390831469) funded by the German Research Foundation (DFG). UR would like to acknowledge support from the Department of Atomic Energy, Government of India (Project Identification Number RTI 4002) and from the J.~C.\ Bose Grant of the Anusandhan National Research Foundation (ANRF), Government of India (ANRF/JBG/2025/000265/PS).

%=============================================================================
\appendix
\section{Circuit Implementation}
\label{app:implementation}
%=============================================================================

%-----------------------------------------------------------------------------
\subsection{Conventional implementation}
\label{app:conventional}
%-----------------------------------------------------------------------------

\begin{figure*}
    \centering
    \colorbox{QCcolor}{\includegraphics[width=\linewidth]{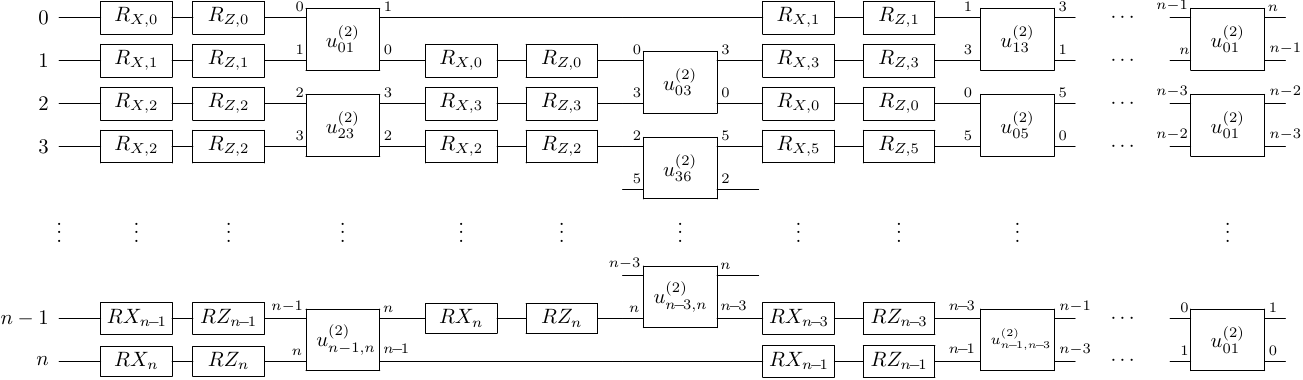}}
    \caption{The quantum circuit corresponding to one second-order Trotter step in the conventional implementation of collective neutrino oscillations. We use the shorthand notation \smash{$R_{X,j} \equiv R_X(\theta=-dt\,\Delta_j\,b_x/(n-1))$} and \smash{$R_{Z,j} \equiv R_Z(\theta=-dt\,\Delta_j\,b_z/(n-1))$}. The composite two-qubit gate $u_{ij}^{(2)}$ carries out the operation \smash{$\exp(-i (dt \, J_{ij} + \frac{\pi}{4}) \hat\sigma_i \otimes \hat\sigma_j)$}. The extra phase $-i \pi/4$ in the \smash{$u_{ij}^{(2)}$} implies that each layer of entangling gates also changes the mapping of logical qubits (small numbers on the wires) onto physical qubits, making the circuit suitable for hardware with limited connectivity at no extra cost.}
    \label{fig:conventional_circuit}
\end{figure*}

In most of the literature on collective neutrino oscillations on a quantum computer, the Hamiltonian that is implemented is the one from \cref{eq:Trotter,eq:hpq}. We show a typical quantum circuit in \cref{fig:conventional_circuit}. It consists of alternating layers of single-qubit gates that implement vacuum oscillations and entangling gates that implement self-interactions. This increases the circuit depth compared to a first-order Trotter approximation where a single set of 1-qubit gates precedes the layers of 2-qubit gates; but since single-qubit gates do not contribute much to the noise, the reduction of the Trotter error is more important here than keeping the number of single-qubit gates minimal.

\begin{figure*}
    \centering
    \colorbox{QCcolor}{\includegraphics[width=\linewidth]{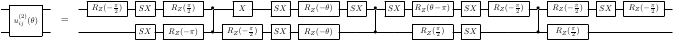}}
    \caption{Implementation of the $u_{ij}^{(2)}$ gate on the IBM Heron QPU.}
    \label{fig:uij}
\end{figure*}

Note that the implementation in \cref{fig:conventional_circuit} only requires nearest-neighbour interactions among qubits. This is because it exploits the fact that $\exp(i \frac{\pi}{4} \hat{\vec\sigma}_p \otimes \hat{\vec\sigma}_q)$ corresponds to a SWAP operation (exchange of quantum states between qubits $p$ and $q$). In other words, if we change $J_{pq} \to J_{pq} + \pi/4$, each application of $h_{pq}$ also changes the mapping between logical and physical qubits. This allows us to implement a full Trotter step, during which each pair of logical qubits interacts once, using $N$ layers of nearest-neighbour entangling gates. At the end of each Trotter step, the ordering of the logical qubits has been exactly reversed (small numbers on the wires in the diagram). Similar approaches to reducing circuit complexity and rendering it suitable for quantum processors with limited connectivity have been described in \cite{Turro:2024shh, Spagnoli:2025etu}.

The decomposition of the $u_{ij}^{(2)}$ gate into native gates depends on the chosen hardware platform. On the IBM Heron QPUs which we have been using, the native 2-qubit gate is $CZ$, and with this in mind we use the implementation shown in \cref{fig:uij}.

%-----------------------------------------------------------------------------
\subsection{Full Dicke approach}
\label{app:full-dicke}
%-----------------------------------------------------------------------------

\paragraph{High-level circuit layout.}
%-------------------------------------
The circuit representing a single Trotter step in the full Dicke approach (\cref{eq:Hs-n-modes}) is shown in \cref{fig:trotter-step-circuit} for a system consisting of two neutrino modes $S_1$, $S_2$, each encoded in a 3-qubit quantum register. The circuit begins with a set of rotations operating on a single register which implement vacuum oscillations, $\exp(-i \sum_j \vec{b}_j \vec{\hat{S}}_j)$. The first two gates labeled $R_{S_z}$ are the generalizations of the standard $R_Z$ gate to Dicke states. They implement the operation \smash{$\exp(-i b_z \hat{S}_z)$}. They are followed by a series of gates of the form \smash{$R_{X}^{(1)}(k,k+1; \theta)$}, which correspond to the operator $\exp(-i (\theta/2) \hat{X}_{k,k+1})$. The superscript $(1)$ indicates that the gate is operating on a single quantum register, $k$, $k+1$ are two of the eight possible states of this register, and the $\hat{X}_{k,k+1}$ operator in the exponent is meant to act selectively between those two states. As we will show below, each $R_{X}^{(1)}(k,k+1; \theta)$ gate requires one (reusable) ancillary qubit. We can already see at this point that some reduction in circuit depth would in principle be possible by providing two ancillary qubits, such that the $R_{X}^{(1)}(k,k+1; \theta)$ operations on the two registers could be carried out simultaneously. We refrain from doing so as our overarching goal in this paper is to keep the number of qubits minimal.

The second part of the Trotter step in \cref{fig:trotter-step-circuit} implements the self-interaction term $\exp(-i \sum_{j<k} J_{jk} \,dt\, \vec{\hat{S}}_j \vec{\hat{S}}_k)$. This term is decomposed according to
\begin{align}
    \exp\Big[ -i \sum_{j<k} J_{jk} \,dt\, \vec{\hat{S}}_j \vec{\hat{S}}_k \Big]
        &= \prod_{j<k} R_{S_zS_z}(j,k; \theta) \, R_X^{(2)}(j,k; \theta)
\end{align}
with $\theta \equiv J_{jk} \, dt$ and 
\begin{align}
    R_{S_zS_z}(j,k;\theta) &\equiv \exp\Big[ -i \theta \,\hat{S}_{j,z} \otimes \hat{S}_{k,z} \Big] \,,
                                                           \label{eq:R-SzSz} \\[0.2cm]
    R_X^{(2)}(j,k; \theta) &\equiv \exp\Big[ -i \tfrac{\theta}{4} \, (\hat{S}^+_j \otimes \hat{S}^-_k
                                                                    + \hat{S}^-_{j+1} \otimes \hat{S}^+_{k-1}) \Big] \,,
                                                           \label{eq:RX2}
\end{align}
where $\hat{S}^\pm_j = \hat{S}_{j,x} \pm i \hat{S}_{j,y}$ are the ladder operators of $su(2)$, restricted to the subspace spanned by the states $\ket{j}$, $\ket{j \pm 1}$ of a quantum register.

\begin{figure*}
    \centering
    \begin{tabularx}{\linewidth}{c@{\qquad}c}
        \colorbox{QCcolor}{\parbox[][3cm][c]{0.17\linewidth}{\includegraphics[width=\linewidth]{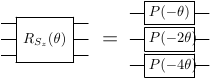}}} &
        \colorbox{QCcolor}{\parbox[][3cm][c]{0.77\linewidth}{\includegraphics[width=\linewidth]{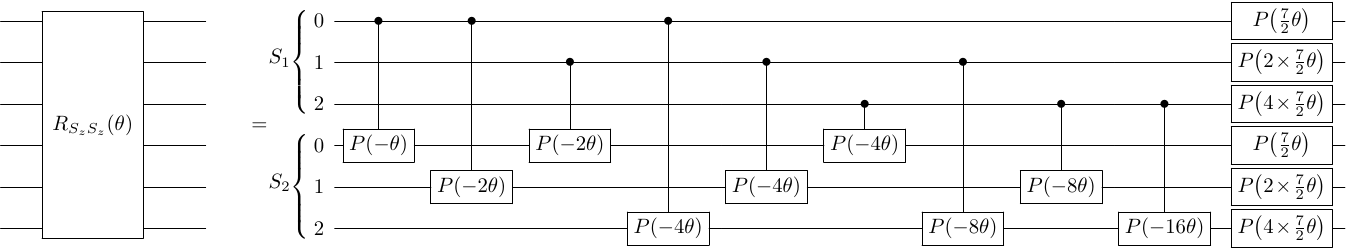}}} \\
        \\[-0.1cm]
        (a) & (b)
    \end{tabularx}
    \caption{The implementation of the quantum gates (a) $R_{S_z}$ and (b) $R_{S_zS_z}$, which correspond to rotations around the $z$-axis on a single 3-qubit Dicke regster or on two such registers, respectively.}
    \label{fig:rsz-gate-circuit}
\end{figure*}

\paragraph{$S_z$ rotations.}
We now describe the detailed implementation of the gates used in \cref{fig:trotter-step-circuit}, beginning in \cref{fig:rsz-gate-circuit}~(a) with the $R_{S_z}$ gate. It is comprised of one phase gate (or, equivalently, $R_Z$ gate) per qubit. The rotation angle in the $p$-th qubit in the register is weighted with the significance $2^p$ of that qubit. Closely related is the two-register $R_{S_zS_z}$ gate, see \cref{eq:R-SzSz} and \cref{fig:rsz-gate-circuit}~(b). Given that the integer value $j$ encoded in one of our registers is related to the $S_z$ quantum number $m$ via $m = j - S$, the the two-register state $\ket{j,k}$ should receive a phase proportional to $(j-S)(k-S) = j k - S j - S k + S^2$. Dropping the global phase $\propto S^2$, we see that we need single-qubit phase gates of the form $P(2^p \theta)$ for the $p$-th qubit, and controlled phase gates between all qubit pairs $(p,q)$ with phase $-2^{p + q} S\,\theta$.

\begin{figure*}
    \centering
    \begin{tabularx}{\linewidth}{c}
        \colorbox{QCcolor}{\parbox[][0.7cm][c]{\linewidth}{}
            \includegraphics[width=\linewidth]{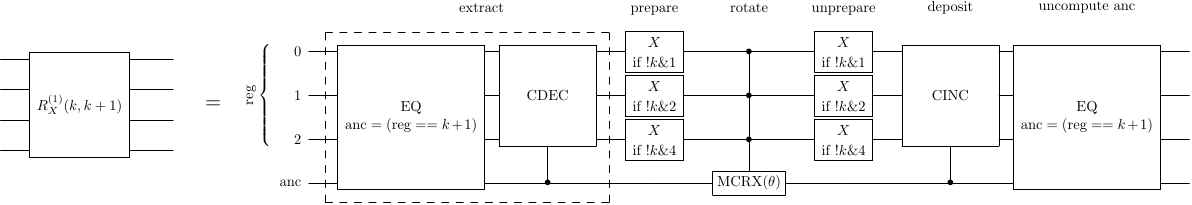}} \\[0.4cm]
        (a) \\[0.4cm]
        \colorbox{QCcolor}{\parbox[][0.7cm][c]{\linewidth}{}
            \includegraphics[width=\linewidth]{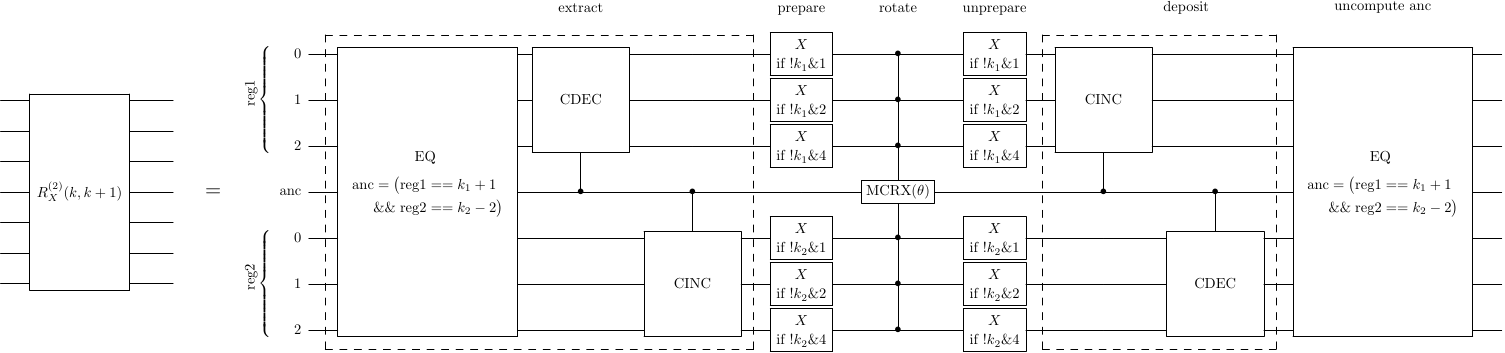}} \\[0.4cm]
        (b)
    \end{tabularx}
    \caption{The ancilla-based implementation of the quantum gates (a) $R_X^{(1)}(k,k+1; \theta)$ and (b) $R_X^{(2)}(j,k; \theta)$ on a single 3-qubit Dicke register or on two such registers, respectively.}
    \label{fig:rx-gate-circuit}
\end{figure*}

\paragraph{$R_X^{(1)}$ and $R_X^{(2)}$ rotations using ancillary qubit.}
%-----------------------------------------------------------------------------
The implementation of the $R_X^{(1)}(k,k+1; \theta)$ and $R_X^{(2)}(j,k; \theta)$ gates is somewhat more involved. \Cref{fig:rx-gate-circuit}~(a) schematically shows the steps needed to perform this operation on a single 3-qubit quantum register. First, the ancillary qubit is flipped from $\ket{0}$ to $\ket{1}$ if the integer encoded in the register is $k+1$. Then, the ancilla is used to control an integer decrement operation on the register. This way, if the register was originally in state $k+1$, it is now in state $k$, and the fact that a decrement has occurred is signalled by the ancilla. Next, those qubits which are zero in state $\ket{k}$ are flipped. Now, if the register was originally in state $\ket{k}$ or $\ket{k+1}$, all qubits are in the $\ket{1}$ state. For any other value of the register, one or several qubits are $\ket{0}$. Therefore, we can now carry out the actual $R_X$ rotation on the ancilla, controlled by all qubits in the register. In other words, the rotation happens only if the register was originally in the $\ket{k}$ or $\ket{k+1}$ state. With the main operation done, we first undo the qubit flips singling out the $\ket{k}$ state. We then apply a controlled increment, conditioned on the -- now rotated -- ancilla. This is when the result of the computation is stored in the register. Finally, the ancilla is uncomputed to return it to zero for its next use. Through this sequence of quantum logic operations, we have effectively carried out an $R_X$ rotation in the $(\ket{k}, \ket{k+1})$ subspace. The two-register version of the gate, shown in \cref{fig:rx-gate-circuit}~(b) follows largely the same logic. The main difference is that the ancilla is now conditioned on the state of both registers, the controlled decrement/increment operations are applied to both registers, and the main rotation is conditioned on the state of both registers.

\begin{figure*}
    \centering
    \begin{tabularx}{\linewidth}{c@{\qquad}c}
        \colorbox{QCcolor}{\parbox[][3cm][c]{0.47\linewidth}{\includegraphics[width=\linewidth]{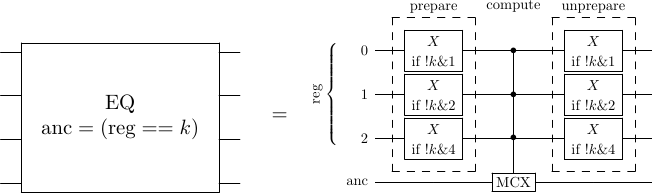}}} &
        \colorbox{QCcolor}{\parbox[][3cm][c]{0.47\linewidth}{\includegraphics[width=\linewidth]{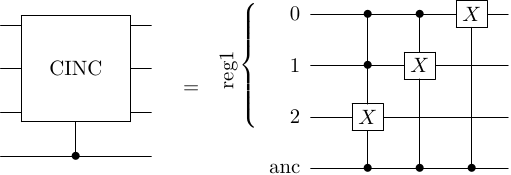}}} \\
        \\
        \multicolumn{2}{c}{
            \colorbox{QCcolor}{\parbox[][4.4cm][c]{0.77\linewidth}{\includegraphics[width=\linewidth]{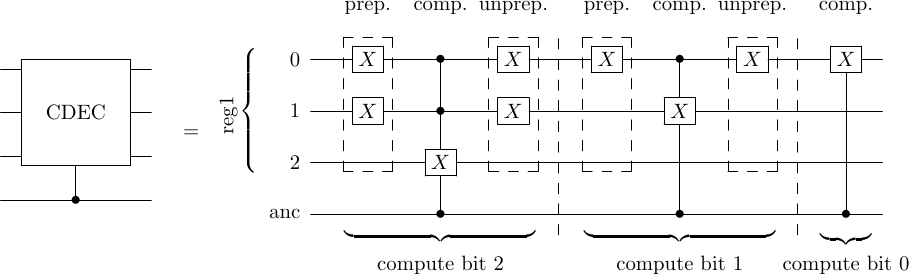}}}
        }
    \end{tabularx}
    \caption{The implementation of the quantum gates (a) EQ (b) controlled increment, and (c) controlled decrement on a 3-qubit quantum register. The implementation of the two-qubit equality gate used in the circuit for \smash{$R_X^{(2)}(j,k; \theta)$} (see \cref{fig:rx-gate-circuit}~(b)) is fully analogous to the single-register version.}
    \label{fig:eq-cint-cdec-circuits}
\end{figure*}

In \cref{fig:eq-cint-cdec-circuits}, we finally describe the atomic operations entering the circuits in \cref{fig:rx-gate-circuit}. The equality gate, EQ, first flips all qubits which in the desired state $\ket{k}$ are zero. Then, the full register is used to control a flip of the ancillary qubit, whereby the latter is transferred to the $\ket{1}$ state if and only if the register was originally in state $\ket{k}$. The register is then returned to its original state.

A conditional increment is implemented by applying a series of conditioned bit flips to the register, starting from the most significant qubit. Each bit flip ($X$ gate) is conditioned on the value of the ancilla and on the values of all lower-significance qubits. The controlled decrement is similar, but here the lower-signifiance qubits are flipped before being used to condition an $X$ gate. That way, any given qubit is flipped only if all lower-significance qubits were originally zero, as is appropriate for a binary decrement.

\begin{figure*}
    \centering
    \begin{tabularx}{1.0\linewidth}{c}
        \colorbox{QCcolor}{\parbox[][0.8cm][c]{\linewidth}{}
            \includegraphics[width=\linewidth]{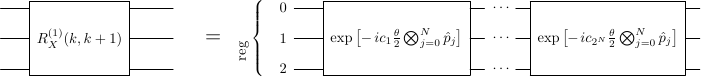}} \\[0.7cm]
        \colorbox{QCcolor}{\parbox[][0.8cm][c]{\linewidth}{}
            \includegraphics[width=\linewidth]{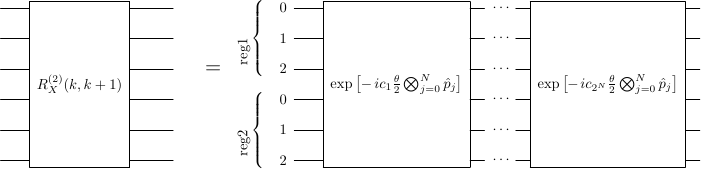}}
    \end{tabularx}
    \caption{Ancilla-free implementation of the quantum gates (a) $R_X^{(1)}(k,k+1; \theta)$ and (b) $R_X^{(2)}(j,k; \theta)$ on a single 3-qubit Dicke register or on two such registers, respectively.}
    \label{fig:rx-gate-circuit-alt}
\end{figure*}

\begin{figure*}
    \centering
    \colorbox{QCcolor}{\parbox[][3cm][c]{\linewidth}{\includegraphics[width=\linewidth]{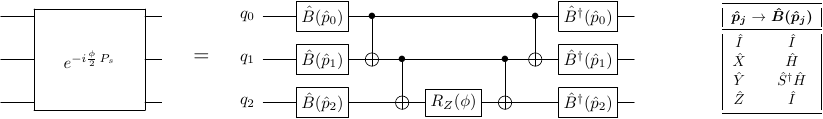}}}
    \caption{Implementation of a Pauli string rotation (\cref{eq:pauli-string-rotation}). First, a set of single-qubit operations transforms each qubit to the eigenbasis of the corresponding Pauli operator \smash{$\hat{p}_j \in \{ \hat{I}, \hat{X}, \hat{Y}, \hat{Z} \}$}. A CNOT cascade then computes the parity of all active qubits into a single qubit (here \#2), which then receives an $R_Z$ rotation. Finally, the CNOT cascade and basis change are reversed. The table on the right lists the basis transformation $\hat{B}(\hat{p}_j)$ that should be applied to the $j$-th qubit, depending on which ``letter'' $\hat{p}_j$ corresponds to in the Pauli string $P_s$.}
    \label{fig:pauli-rotation}
\end{figure*}

\paragraph{$R_X^{(1)}$ and $R_X^{(2)}$ rotations -- ancilla-free implementation.}
%----------------------------------------------------------------------------------
In addition to the $\smash{R_X^{(1)}}$ and $\smash{R_X^{(2)}}$ gate implementations shown in \cref{fig:rx-gate-circuit}, we have also developed an alternative, ancilla-free implementation, which we show in \cref{fig:rx-gate-circuit-alt,fig:pauli-rotation}. The ancilla-free implementation tends to lead to deeper circuits, but it is truly minimal in terms of the number of qubits used. For \smash{$R_X^{(1)}$}, we write
\begin{align}
    R_{X}^{(1)}(k,k\!+\!1; \theta)
        &= \exp[-i (\theta/2) \hat{X}_{k,k+1}] \notag\\[0.2cm]
        &= \exp\Big[ -i \tfrac{\theta}{2} \, \Big( \ket{k+1}\!\bra{k}
           + \ket{k}\!\bra{k+1} \Big) \Big] ,
    \label{eq:RX1-ancilla-free}
\end{align}
and then decompose the outer product $\ket{k+1}\!\bra{k}$ and its hermitian conjugate into Pauli strings according to
\begin{align}
    \ket{k+1}\!\bra{k} = \bigotimes_{j=0}^{N} \hat{o}_j \,.
    \label{eq:transition-factored}
\end{align}
Here, $\hat{o}_j$ is an operation applied to the $j$-th qubit, and $\otimes$ denotes the Kronecker product. Each $\hat{o}_j$ is one of the following four operators:
\begin{align}
    \begin{split}
        0 \!\to\! 0 \;&:\quad
            \hat{o}_j = \ket{0}\!\bra{0}
            = \tfrac{1}{2}\big(\hat{I} + \hat{Z}_j\big), \\[4pt]
        1 \!\to\! 1 \;&:\quad
            \hat{o}_j = \ket{1}\!\bra{1}
            = \tfrac{1}{2}\big(\hat{I} - \hat{Z}_j\big), \\[4pt]
        0 \!\to\! 1 \;&:\quad
            \hat{o}_j = \ket{1}\!\bra{0}
            = \tfrac{1}{2}(\hat{X}_j + i\hat{Y}_j), \\[4pt]
        1 \!\to\! 0 \;&:\quad
            \hat{o}_j = \ket{0}\!\bra{1}
            = \tfrac{1}{2}(\hat{X}_j - i\hat{Y}_j).
    \end{split}
    \label{eq:local-pauli-factors}
\end{align}
The hermitian conjugate outer product $\ket{k}\!\bra{k+1}$ can be decomposed in a similar way. Combining the two and expanding the tensor product in \cref{eq:transition-factored} yields up to $2^N$ terms, each one a Kronecker product of elementary $\hat{I}$, $\hat{X}$, $\hat{Y}$, and $\hat{Z}$ operations (a ``Pauli string''):
\begin{align}
    \ket{k+1}\!\bra{k1} + \text{h.c.}
        = \sum_{s} 2\,\text{Re}(c_s) \; \hat{P}_s \,,
    \label{eq:pauli-string-sum}
\end{align}
with $\hat{P}_s$ composed of single qubit $\{\hat{I}, \hat{X}, \hat{Y}, \hat{Z}\}$ and with numerical coefficients $c_s$.

The task is now to evaluate a series of operations of the form
\begin{align}
    \exp\Big[
        -i c_s \tfrac{\theta}{2} \, \bigotimes\nolimits_{j=0}^N \hat{p}_j \,,
    \Big]
    \label{eq:pauli-string-rotation}
\end{align}
with $\hat{p}_j \in \{ \hat{I}, \hat{X}, \hat{Y}, \hat{Z} \}$. A general method for implementing such a multi-qubit gate is the following: first, each qubit is transformed to the eigenbasis of the corresponding $\hat{p}_j$ operator. If $\hat{p}_j = \hat{X}$, this is achieved by applying a Hadamard ($\hat{H}$) gate. For $\hat{p}_j = \hat{Y}$, the appropriate operation is $\hat{S}^\dag \hat{H}$, where $\hat{S}^\dag = R_Z(-\pi/2)$. If $\hat{p}_j = \hat{I}, \hat{Z}$, we are already in the correct basis. The qubits are then entangled via a CNOT chain and an $R_Z(c_s \theta)$ rotation is applied to the last qubit in the chain. Finally, the entanglement and basis change are undone via a reversed CNOT chain, followed by the inverse basis change operations on each qubit.

The implementation of the $R_X^{(2)}$ gate follows an identical logic. For a rotation in the $(\ket{j; k}, \ket{j+1; k-1})$ subspace of the two quantum registers of size $N_1$ and $N_2$, the algorithm determines the combined bit strings of length $N_1+N_2$ corresponding to the $\ket{j; k}$ and $\ket{j+1; k-1}$ states. By comparing these two bit strings, it determines which of the four operations $\ket{0} \!\to\! \ket{0}$, $\ket{0} \!\to\! \ket{1}$, $\ket{1} \!\to\! \ket{0}$, $\ket{1} \!\to\! \ket{1}$ each qubit should undergo. Based on this, it constructs the appropriate Pauli strings and then applies them in the same way as before via a set of basis rotations, a CNOT chain (across both registers), an $R_Z$ rotation, an inverse CNOT chain, and the inverse basis transformations.

\paragraph{Example: two-qubit register ($k=2$).}
%----------------------------------------------
To make the ancilla-free implementation of the \smash{$R_X^{(1)}$} gate more transparent, we now discuss as a concrete example: the transition $\ket{01} \to \ket{10}$ in a two-qubit register. Following \cref{eq:local-pauli-factors}, this corresponds to
\begin{align}
    \hat{o}_0 &= \ket{0}\!\bra{1}
              = \tfrac{1}{2}(\hat{X}_0 - i\hat{Y}_0), \\[4pt]
    \hat{o}_1 &= \ket{1}\!\bra{0}
              = \tfrac{1}{2}(\hat{X}_1 + i\hat{Y}_1).
\end{align}
The Kronecker product $\hat{o}_1\otimes\hat{o}_0$ expands to four Pauli strings, namely
\begin{align}
    \ket{10}\!\bra{01}
    \;=\;
    \tfrac{1}{4}\Bigl[
        \hat{X}_1\hat{X}_0
        \;-\; i\,\hat{X}_1\hat{Y}_0
        \;+\; i\,\hat{Y}_1\hat{X}_0
        \;+\; \hat{Y}_1\hat{Y}_0
    \Bigr].
\end{align}
However, when adding the Hermitian conjugate, the second and third term cancel, leaving only
\begin{align}
    \ket{10}\!\bra{01} + \mathrm{h.c.}
    \;=\;
    \tfrac{1}{2}\!\left(
        \hat{X}_1\hat{X}_0 + \hat{Y}_1\hat{Y}_0
    \right).
    \label{eq:2qubit-flip-pauli}
\end{align}
We therefore need to carry out only two Pauli-string rotations:
\begin{align}
    \exp\!\bigl[-i\,\theta\,\big(\ket{10}\!\bra{01}+\mathrm{h.c.} \big)\,\bigr]
    \;\approx\;&
    \exp\!\Bigl[-i\,\tfrac{\theta}{2}\;\hat{X}_1\hat{X}_0\Bigr]
    \notag\\&\times
    \exp\!\Bigl[-i\,\tfrac{\theta}{2}\;\hat{Y}_1\hat{Y}_0\Bigr].
    \label{eq:flip-exp-sum}
\end{align}

%-----------------------------------------------------------------------------
\subsection{Dicke approach on a diagonal subspace.}
\label{app:diagonal-dicke}
%-----------------------------------------------------------------------------

\begin{figure*}
    \centering
    \colorbox{QCcolor}{\parbox[][3cm][c]{\linewidth}{\includegraphics[width=\linewidth]{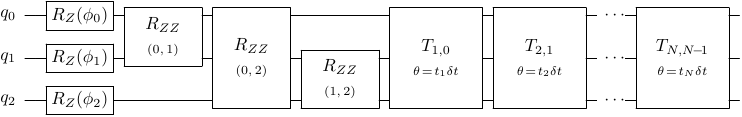}}}
    \caption{Quantum circuit for a single Trotter step in the diagonal Dicke approach, illustrated here for a 3-qubit quantum register. The diagonal evolution $e^{-i\hat{H}_D\,\delta t/2}$ (\cref{eq:HD-polynomial}) consists of single-qubit $R_Z$ rotations from the linear term $\beta\hat{n}$ and two-qubit $R_{ZZ}$ gates from the quadratic term $\alpha\hat{n}^2$, acting on all $\binom{k}{2}$ qubit pairs. The off-diagonal evolution $e^{-i\hat{H}_T\,\delta t}$ (\cref{eq:HT-offdiag}) is decomposed into blocks $T_{i,i-1}$, one for each transition $\ket{i}\!\bra{i\!-\!1}$ and $\ket{i\!-\!1}\!\bra{i}$\ with coupling strength $t_i$ (\cref{eq:ti-couplings}). Each block is further decomposed into Pauli string rotations as described in \cref{fig:pauli-rotation}.}
    \label{fig:diag-circuit-block}
\end{figure*}

We now turn to the Dicke approach for a system with two neutrino populations, each consisting of $N$ particles, but restricted to the ``diagonal'' Hilbert space from \cref{eq:H_diag,eq:H_offdiag}. This diagonal subspace can be encoded on $k = \lceil \log_2(N+1)\rceil$ qubits using the computational-basis mapping
\begin{align}
    \ket{i}_{\text{comp}}
        \;\longleftrightarrow\;
        \ket{m = i - S,\; -(i-S)},
    \label{eq:diag-encoding}
\end{align}
with $i \in \{0,1,\ldots,N\}$ and $S = N/2$. In other words, we again encode the $N+1$ Dicke states as binary numbers stored in a quantum register. Initial states are prepared by applying Pauli $X$ gates to those qubits whose corresponding bits in the binary representation are 1. When $2^k > N+1$, the computational basis states $\ket{N+1}$ through $\ket{2^k - 1}$ are unphysical; these require special treatment, which we will discuss below. The most efficient use of computational resources is typically achieved when $N+1$ is a power of two, so that all states available in the quantum register are used to encode physical information.

\paragraph{Hamiltonian decomposition.}
%-------------------------------------
Recalling \cref{eq:H_diag,eq:H_offdiag}, the subspace Hamiltonian takes on a tridiagonal form and can be split into a diagonal part $\hat{H}_D$ and an off-diagonal part $\hat{H}_T$:
\begin{align}
    \hat{H} = \hat{H}_D + \hat{H}_T \,.
    \label{eq:diag-H-split}
\end{align}
The diagonal part is a polynomial in the number operator $\hat{n} \equiv \sum_{j=0}^{k-1} 2^j \ket{1}_j\!\bra{1}_j$, as can be seen by substituting $m = \hat{n} - S$ in \cref{eq:H_diag}. The result can be written as:
\begin{align}
    \hat{H}_D &= \alpha\,\hat{n}^2 + \beta\,\hat{n} + \gamma\,\mathbbm{1} \,,
    \label{eq:HD-polynomial}
\end{align}
with the coefficients
\begin{align}
    \begin{split}
        \alpha &= -4J, \\[0.1cm]
        \beta  &= -2\Delta\cos 2\theta + 8JS, \\[0.03cm]
        \gamma &= 2\Delta\cos 2\theta\; S - 4J S^2 \,.
    \end{split}
    \label{eq:HD-coefficients}
\end{align}
The off-diagonal part of the Hamiltonian couples adjacent computational basis states:
\begin{align}
    \hat{H}_T = \sum_{i=1}^{N} t_i \big(\ket{i}\!\bra{i-1} + \ket{i-1}\!\bra{i}\big),
    \label{eq:HT-offdiag}
\end{align}
where
\begin{align}
    t_i = 2J\, i\,(N - i + 1)
    \label{eq:ti-couplings}
\end{align}
follows directly from \cref{eq:H_offdiag}, with the same substitution of $m = i - S$. In the following, we discuss the implementation of the Hamiltonian as quantum gates.

\paragraph{Diagonal terms: $\exp(-i \hat{H}_D\,\delta t)$.}
%----------------------------------------------------------

We write the number operator as
\begin{align}
    \hat{n} &\equiv \sum_{j=0}^{k-1} 2^j \, \ket{1}_j\!\bra{1}_j \notag\\
            &=      \sum_{j=0}^{k-1} 2^{j-1} \, (\hat{I} - \hat{Z}_j) \notag\\
            &=      \frac{2^k - 1}{2}\,\mathbbm{1}
                  - \sum_{j=0}^{k-1} 2^{j-1}\, \hat{Z}_j\,,
    \label{eq:nhat-projector}
\end{align}
where $\hat{I}$, $\hat{Z}_j$ are the identity and Pauli-$Z$ operators acting on the $j$-th qubit. For the decomposition of $\hat{n}^2$, we start from the second line of \cref{eq:nhat-projector}. Defining $a_j \equiv 2^{j-1}$, we can write
\begin{align}
    \hat{n}^2
    &= \sum_j a_j^2\,(I - \hat{Z}_j)^2
     + 2\!\sum_{j<l} a_j\, a_l\,(I - \hat{Z}_j)(I - \hat{Z}_l) \,.
    \label{eq:nsq-expand}
\end{align}
Using $(I - \hat{Z}_j)^2 = 2(I - \hat{Z}_j)$, expanding, and collecting terms then gives the expression
\begin{align}
    \hat{n}^2
    &= \Bigl(\sum_j a_j\Bigr)^{\!2}\,\mathbbm{1}
     \,-\, 2\Bigl(\sum_l a_l\Bigr)\!\sum_{j}\! a_j \hat{Z}_j
     \,+\, 2\!\sum_{j<l}\! a_j a_l\,\hat{Z}_j \hat{Z}_l \,.
    \label{eq:nsq-pauli}
\end{align}
Combining \cref{eq:HD-polynomial,eq:nhat-projector,eq:nsq-pauli}, the diagonal evolution operator \smash{$\exp(-i \hat{H}_D\,\delta t)$} decomposes into the following operators, as shown in \cref{fig:diag-circuit-block}:
\begin{enumerate}
    \item \textbf{Single-qubit $R_Z$ gates}: A rotation $R_Z(\phi_j)$ is applied to each qubit, where the rotation angle for the $j$-th qubit is \smash{$\phi_j = -2 \big( -2^{j-1} \beta + \alpha\, c_j^{(1)} \big)\,\delta t$}, where
    \begin{align}
        c_j^{(1)} \equiv -2 \Bigl(\sum_l a_l\Bigr) \! \sum_{j} a_j \,.
    \end{align}
    This is completely analogous to the circuit shown above in \cref{fig:rsz-gate-circuit}\,(a) for the specific example of a 3-qubit register.

    \item \textbf{Two-qubit $R_{ZZ}$ gates}: For each pair $(j,l)$ with $j < l$, a rotation \smash{$R_{ZZ}(\phi_{jl}) \equiv \exp(-i\frac{\phi_{jl}}{2} \hat{Z}_j \otimes \hat{Z}_l)$} is applied, with \smash{$\phi_{jl} = -2\alpha\, c_{jl}^{(2)}\,\delta t$}, where
    \begin{align}
        c_{jl}^{(2)} \equiv 2 \sum_{j<l} a_j \, a_l \,.
    \end{align}
    Each $R_{ZZ}$ gate is implemented using two CNOT gates and a single $R_Z$ rotation.
\end{enumerate}
We neglect global phases as they do not affect measurable quantities.

\paragraph{Off-diagonal evolution: $\exp(-i \hat{H}_T\,\delta t)$.}
%------------------------------------------------------------------

The off-diagonal part of the Hamiltonian, \cref{eq:HT-offdiag}, is a sum of two-level transition operators between  adjacent computational basis states $\ket{i-1}$ and $\ket{i}$. It is effectively a local Pauli $X$ gate in the subspace spanned by these two states. Therefore, it is structurally identical to the $\smash{R_X^{(1)}}$ gate from \cref{app:full-dicke} and can be implemented in the same way. We choose here the ancilla-free implementation depicted in \cref{fig:rx-gate-circuit-alt,fig:pauli-rotation}.

\begin{figure*}
    \centering
    \colorbox{QCcolor}{\includegraphics[width=\linewidth]{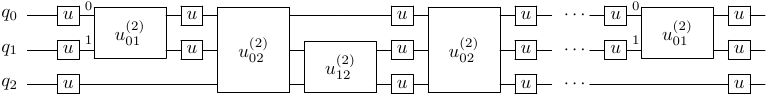}}
    \caption{KAK decomposed quantum circuit for a single Trotter step in the diagonal Dicke approach. We show the general layout of the circuit for a three-qubit simulation that represents 14 total physical neutrinos divided into 2 modes of equal number. We leave at abstract two qubit and single qubit gates level since these are artifacts of KAK decomposition that does not directly hold physical meaning. }
    \label{fig:kak-circuit-block}
\end{figure*}

%=============================================================================
\section{Noise Mitigation}
\label{app:error_mitigation}
%=============================================================================

All quantum simulations shown in this paper have been carried out on QPU of the IBM Heron~r3 generation, namely the IBM Boston device. For access to IBM's quantum ecosystem, we have used the Qiskit Estimator V2 primitives \cite{javadiabhari2024quantumcomputingqiskit}. We set \texttt{resilience\_level = 2} in Qiskit to enable the following noise mitigation techniques:

\paragraph{Dynamical Decoupling.}
Quantum circuits on superconducting hardware suffer from decoherence error if a qubit stays idle for an extended period of time. This is due to natural thermal fluctuations, environmental noise, and other sources of noise. Dynamical decoupling \cite{PhysRevLett.82.2417} mitigates this problem by applying dummy pulse sequences to idle qubits, which are designed such that the time-integrated interaction Hamiltonian between the qubit and its thermal environment averages to zero in the limit that the pulse sequence is much faster than the decoherence time \cite{javadiabhari2024quantumcomputingqiskit}. We use the $X_pX_m$ sequence.

\paragraph{Zero-Noise Extrapolation.}
The goal of this technique is to characterize the dependence of the circuit output on the noise level and to then extrapolate to zero noise. To this end, the circuit is run not only at the baseline noise level, but also with added noise, implemented in the form of dummy gate sequences (gate sequences that are mathematically equivalent to the identity, but increase the circuit depth and therefore the noise level). The circuit observables are measured at noise amplification factors of $\{1, 2, 3\}$ and then extrapolated with a linear model to the theoretical zero-noise limit to recover the ideal result \cite{Temme:2016vkz, 9259940}.

\paragraph{Pauli Twirling.}
Systematic errors such as gate over-rotations can hinder the assumptions underlying the zero-noise extrapolation and disrupt its efficiency. Pauli Twirling mitigates this problem by inserting random low-noise single-qubit Pauli gates around noisy entangling gates. While the added gates are chosen such that, mathematically, they do not alter the circuit, their presence changes coherent noise channels to stochastic ones. The latter can be mitigated in combination with the other techniques discussed here \cite{Wallman:2015uzh}.

\paragraph{Twirled Readout Error Extinction.}
To address errors during circuit readout, we utilize Twirled Readout Error Extinction (TREX) \cite{PhysRevA.105.032620}. This is similar to Pauli twirling, but applied to the measurement process rather than internal entangling gates. TREX effectively diagonalizes the readout error matrix and allows error mitigation through measurement of these diagonal factors instead of the full matrix.

\paragraph{Default Run Configuration.}
\Cref{tab:runtime_params} summarizes the default runtime parameters for the hardware runs in this work. Any modifications to this set of parameters tailored to specific simulations were mentioned explicitly. 

\begin{table}[h]
    \begin{tabular}{l@{\qquad}l}
        \toprule
        \textbf{Parameter} & \textbf{Value} \\
        \midrule
        Resilience level & 2 \\
        ZNE noise factors & 1, 2, 3 \\
        ZNE extrapolator & Linear \\
        Dynamical decoupling & Enabled ($X_pX_m$) \\
        Shots per circuit & \num{1\,024} \\
        \bottomrule
    \end{tabular}
    \centering
    \caption{The default error mitigation parameters used in this work.}
`   \label{tab:runtime_params}
\end{table}

%=============================================================================
\bibliography{refs.bib}

@PREAMBLE{
 "\providecommand{\noopsort}[1]{}" 
 # "\providecommand{\singleletter}[1]{#1}%" 
}

@article{KamLAND:2002uet,
    author = "Eguchi, K. and others",
    collaboration = "KamLAND",
    title = "{First results from KamLAND: Evidence for reactor anti-neutrino disappearance}",
    eprint = "hep-ex/0212021",
    archivePrefix = "arXiv",
    doi = "10.1103/PhysRevLett.90.021802",
    journal = "Phys. Rev. Lett.",
    volume = "90",
    pages = "021802",
    year = "2003"
}

@article{MINOS:2011amj,
    author = "Adamson, P. and others",
    collaboration = "MINOS",
    title = "{Improved search for muon-neutrino to electron-neutrino oscillations in MINOS}",
    eprint = "1108.0015",
    archivePrefix = "arXiv",
    primaryClass = "hep-ex",
    reportNumber = "FERMILAB-PUB-11-351-PPD, BNL-96120-2011-JA",
    doi = "10.1103/PhysRevLett.107.181802",
    journal = "Phys. Rev. Lett.",
    volume = "107",
    pages = "181802",
    year = "2011"
}

@article{Pantaleone:1992eq,
    author = "Pantaleone, James T.",
    title = "{Neutrino oscillations at high densities}",
    reportNumber = "DOE-ER-40561-056, INT-92-07-01",
    doi = "10.1016/0370-2693(92)91887-F",
    journal = "Phys. Lett. B",
    volume = "287",
    pages = "128--132",
    year = "1992"
}

@article{Pantaleone:1992xh,
    author = "Pantaleone, James T.",
    title = "{Dirac neutrinos in dense matter}",
    reportNumber = "FERMILAB-PUB-92-018-T, UCRHEP-T84",
    doi = "10.1103/PhysRevD.46.510",
    journal = "Phys. Rev. D",
    volume = "46",
    pages = "510--523",
    year = "1992"
}

@article{Qian:1994wh,
    author = "Qian, Yong Zhong and Fuller, George M.",
    title = "{Neutrino-neutrino scattering and matter enhanced neutrino flavor transformation in Supernovae}",
    eprint = "astro-ph/9406073",
    archivePrefix = "arXiv",
    reportNumber = "DOE-ER-40561-150, INT-94-00-63",
    doi = "10.1103/PhysRevD.51.1479",
    journal = "Phys. Rev. D",
    volume = "51",
    pages = "1479--1494",
    year = "1995"
}

@article{Samuel:1995ri,
    author = "Samuel, Stuart",
    title = "{Bimodal coherence in dense selfinteracting neutrino gases}",
    eprint = "hep-ph/9604341",
    archivePrefix = "arXiv",
    reportNumber = "MPI-PHT-95-57, CCNY-HEP-95-5",
    doi = "10.1103/PhysRevD.53.5382",
    journal = "Phys. Rev. D",
    volume = "53",
    pages = "5382--5393",
    year = "1996"
}

@article{Pastor:2002we,
    author = "Pastor, Sergio and Raffelt, Georg",
    title = "{Flavor oscillations in the supernova hot bubble region: Nonlinear effects of neutrino background}",
    eprint = "astro-ph/0207281",
    archivePrefix = "arXiv",
    reportNumber = "MPI-PHT-2002-26",
    doi = "10.1103/PhysRevLett.89.191101",
    journal = "Phys. Rev. Lett.",
    volume = "89",
    pages = "191101",
    year = "2002"
}

@article{Balantekin:2004ug,
    author = "Balantekin, A. B. and Yuksel, H.",
    title = "{Neutrino mixing and nucleosynthesis in core-collapse supernovae}",
    eprint = "astro-ph/0411159",
    archivePrefix = "arXiv",
    doi = "10.1088/1367-2630/7/1/051",
    journal = "New J. Phys.",
    volume = "7",
    pages = "51",
    year = "2005"
}

@article{Spagnoli:2025etu,
    author = "Spagnoli, Luca and others",
    title = "{Collective neutrino oscillations in three flavors on qubit and qutrit processors}",
    eprint = "2503.00607",
    archivePrefix = "arXiv",
    primaryClass = "quant-ph",
    reportNumber = "N3AS-24-010, RIKEN-iTHEMS-Report-25, NT@UW-25-4",
    doi = "10.1103/gjr1-lf8s",
    journal = "Phys. Rev. D",
    volume = "111",
    number = "10",
    pages = "103054",
    year = "2025"
}

@article{art:AtmoNuSuper-Kamiokande:1998kpq,
    author = "Fukuda, Y. and others",
    collaboration = "Super-Kamiokande",
    title = "{Evidence for oscillation of atmospheric neutrinos}",
    eprint = "hep-ex/9807003",
    archivePrefix = "arXiv",
    reportNumber = "BU-98-17, ICRR-REPORT-422-98-18, UCI-98-8, KEK-PREPRINT-98-95, LSU-HEPA-5-98, UMD-98-003, SBHEP-98-5, TKU-PAP-98-06, TIT-HPE-98-09",
    doi = "10.1103/PhysRevLett.81.1562",
    journal = "Phys. Rev. Lett.",
    volume = "81",
    pages = "1562--1567",
    year = "1998"
}

@article{art:SolarNeutrino,
    author = {Nakahata, Masayuki},
    title = {History of solar neutrino observations},
    journal = {Progress of Theoretical and Experimental Physics},
    volume = {2022},
    number = {12},
    pages = {12B103},
    year = {2022},
    month = {03},
    issn = {2050-3911},
    doi = {10.1093/ptep/ptac039},
    url = {https://doi.org/10.1093/ptep/ptac039},
    eprint = {https://academic.oup.com/ptep/article-pdf/2022/12/12B103/48422428/ptac039.pdf},
}

@article{Hall:2021rbv,
    author = "Hall, Benjamin and Roggero, Alessandro and Baroni, Alessandro and Carlson, Joseph",
    title = "{Simulation of collective neutrino oscillations on a quantum computer}",
    eprint = "2102.12556",
    archivePrefix = "arXiv",
    primaryClass = "quant-ph",
    doi = "10.1103/PhysRevD.104.063009",
    journal = "Phys. Rev. D",
    volume = "104",
    number = "6",
    pages = "063009",
    year = "2021"
}

@article{Amitrano:2022yyn,
    author = "Amitrano, Valentina and Roggero, Alessandro and Luchi, Piero and Turro, Francesco and Vespucci, Luca and Pederiva, Francesco",
    title = "{Trapped-ion quantum simulation of collective neutrino oscillations}",
    eprint = "2207.03189",
    archivePrefix = "arXiv",
    primaryClass = "quant-ph",
    doi = "10.1103/PhysRevD.107.023007",
    journal = "Phys. Rev. D",
    volume = "107",
    number = "2",
    pages = "023007",
    year = "2023"
}

@article{Siwach:2023wzy,
    author = "Siwach, Pooja and Harrison, Kaytlin and Balantekin, A. Baha",
    title = "{Collective neutrino oscillations on a quantum computer with hybrid quantum-classical algorithm}",
    eprint = "2308.09123",
    archivePrefix = "arXiv",
    primaryClass = "quant-ph",
    reportNumber = "LLNL-JRNL-853275-DRAFT",
    doi = "10.1103/PhysRevD.108.083039",
    journal = "Phys. Rev. D",
    volume = "108",
    number = "8",
    pages = "083039",
    year = "2023"
}

@article{Tripathi:2025cok,
    author = "Tripathi, Shvetaank and Joshi, Sandeep and Rajpoot, Garima and Shukla, Prashant",
    title = "{Quantum Simulation of Collective Neutrino Oscillations in Dense Neutrino Environment}",
    eprint = "2508.11610",
    archivePrefix = "arXiv",
    primaryClass = "quant-ph",
    month = "8",
    year = "2025",
    journal = "",
}

@article{Shalgar:2023ooi,
    author = "Shalgar, Shashank and Tamborra, Irene",
    title = "{Do we have enough evidence to invalidate the mean-field approximation adopted to model collective neutrino oscillations?}",
    eprint = "2304.13050",
    archivePrefix = "arXiv",
    primaryClass = "astro-ph.HE",
    doi = "10.1103/PhysRevD.107.123004",
    journal = "Phys. Rev. D",
    volume = "107",
    number = "12",
    pages = "123004",
    year = "2023"
}

@article{Friedland:2003dv,
    author = "Friedland, Alexander and Lunardini, Cecilia",
    title = "{Neutrino flavor conversion in a neutrino background: Single particle versus multiparticle description}",
    eprint = "hep-ph/0304055",
    archivePrefix = "arXiv",
    reportNumber = "LA-UR-03-2162, NSF-KITP-03-20",
    doi = "10.1103/PhysRevD.68.013007",
    journal = "Phys. Rev. D",
    volume = "68",
    pages = "013007",
    year = "2003"
}

@article{Friedland:2003eh,
    author = "Friedland, Alexander and Lunardini, Cecilia",
    title = "{Do many particle neutrino interactions cause a novel coherent effect?}",
    eprint = "hep-ph/0307140",
    archivePrefix = "arXiv",
    reportNumber = "LA-UR-03-3847",
    doi = "10.1088/1126-6708/2003/10/043",
    journal = "JHEP",
    volume = "10",
    pages = "043",
    year = "2003"
}

@article{Mikheyev:1985zog,
    author = "Mikheyev, S. P. and Smirnov, A. Yu.",
    title = "{Resonance Amplification of Oscillations in Matter and Spectroscopy of Solar Neutrinos}",
    journal = "Sov. J. Nucl. Phys.",
    volume = "42",
    pages = "913--917",
    year = "1985"
}

@article{Wolfenstein:1977ue,
      author         = "Wolfenstein, L.",
      title          = "{Neutrino Oscillations in Matter}",
      journal        = "Phys.Rev.",
      volume         = "D17",
      pages          = "2369-2374",
      doi            = "10.1103/PhysRevD.17.2369",
      year           = "1978",
      reportNumber   = "COO-3066-102",
      SLACcitation   = "%%CITATION = PHRVA,D17,2369;%%",
      memo      = "Original paper on MSW effect. Discusses oscillations due to
       nonzero neutrino masses as well as due to non-diagonal weak interactions.
       As applications, solar neutrinos and a long baseline experiment (L >= 1000 km)
       are discussed.",
}

@article{Sigl:1993ctk,
    author = "Sigl, G. and Raffelt, G.",
    title = "{General kinetic description of relativistic mixed neutrinos}",
    reportNumber = "MPI-PH-92-112",
    doi = "10.1016/0550-3213(93)90175-O",
    journal = "Nucl. Phys. B",
    volume = "406",
    pages = "423--451",
    year = "1993"
}

@article{Cervia:2019res,
    author = "Cervia, Michael J. and Patwardhan, Amol V. and Balantekin, A. B. and Coppersmith, {\textdaggerdbl} S. N. and Johnson, Calvin W.",
    title = "{Entanglement and collective flavor oscillations in a dense neutrino gas}",
    eprint = "1908.03511",
    archivePrefix = "arXiv",
    primaryClass = "hep-ph",
    doi = "10.1103/PhysRevD.100.083001",
    journal = "Phys. Rev. D",
    volume = "100",
    number = "8",
    pages = "083001",
    year = "2019"
}

@article{Balantekin:2023qvm,
    author = "Balantekin, A. B. and Cervia, Michael J. and Patwardhan, Amol V. and Rrapaj, Ermal and Siwach, Pooja",
    title = "{Quantum information and quantum simulation of neutrino physics}",
    eprint = "2305.01150",
    archivePrefix = "arXiv",
    primaryClass = "nucl-th",
    doi = "10.1140/epja/s10050-023-01092-7",
    journal = "Eur. Phys. J. A",
    volume = "59",
    number = "8",
    pages = "186",
    year = "2023"
}

@article{Yeter-Aydeniz:2021olz,
    author = {Yeter-Aydeniz, K{\"u}bra and Bangar, Shikha and Siopsis, George and Pooser, Raphael C.},
    title = "{Collective neutrino oscillations on a quantum computer}",
    eprint = "2104.03273",
    archivePrefix = "arXiv",
    primaryClass = "quant-ph",
    doi = "10.1007/s11128-021-03348-x",
    journal = "Quant. Inf. Proc.",
    volume = "21",
    number = "3",
    pages = "84",
    year = "2022"
}

@article{Jha:2021itm,
    author = "Jha, Abhishek Kumar and Chatla, Akshay",
    title = "{Quantum studies of neutrinos on IBMQ processors}",
    doi = "10.1140/epjs/s11734-021-00358-9",
    journal = "Eur. Phys. J. ST",
    volume = "231",
    number = "2",
    pages = "141--149",
    year = "2022"
}

@article{Chernyshev:2024pqy,
    author = "Chernyshev, Ivan and Robin, Caroline E. P. and Savage, Martin J.",
    title = "{Quantum magic and computational complexity in the neutrino sector}",
    eprint = "2411.04203",
    archivePrefix = "arXiv",
    primaryClass = "quant-ph",
    reportNumber = "IQuS@UW-21-091",
    doi = "10.1103/PhysRevResearch.7.023228",
    journal = "Phys. Rev. Res.",
    volume = "7",
    number = "2",
    pages = "023228",
    year = "2025"
}

@article{Turro:2024shh,
    author = "Turro, Francesco and Chernyshev, Ivan A. and Bhaskar, Ramya and Illa, Marc",
    title = "{Qutrit and qubit circuits for three-flavor collective neutrino oscillations}",
    eprint = "2407.13914",
    archivePrefix = "arXiv",
    primaryClass = "quant-ph",
    reportNumber = "IQuS@UW-21-082",
    doi = "10.1103/PhysRevD.111.043038",
    journal = "Phys. Rev. D",
    volume = "111",
    number = "4",
    pages = "043038",
    year = "2025"
}

@article{Singh:2024vpu,
    author = "Singh, Gayatri and Arvind and Dorai, Kavita",
    title = "{Simulating three-flavor neutrino oscillations on a nuclear magnetic resonance quantum processor}",
    eprint = "2412.15617",
    archivePrefix = "arXiv",
    primaryClass = "quant-ph",
    doi = "10.1088/1402-4896/adedd4",
    journal = "Phys. Scripta",
    volume = "100",
    number = "8",
    pages = "085106",
    year = "2025"
}

@inproceedings{Akhmedov:1999uz,
    author = "Akhmedov, Evgeny K.",
    title = "{Neutrino physics}",
    booktitle = "{ICTP Summer School in Particle Physics}",
    eprint = "hep-ph/0001264",
    archivePrefix = "arXiv",
    reportNumber = "FISIST-1-2000-CFIF",
    pages = "103--164",
    month = "6",
    year = "1999"
}

@article{Johns:2023ewj,
    author = "Johns, Lucas",
    title = "{Neutrino many-body correlations}",
    eprint = "2305.04916",
    archivePrefix = "arXiv",
    primaryClass = "hep-ph",
    doi = "10.1142/S0217751X24501227",
    journal = "Int. J. Mod. Phys. A",
    volume = "39",
    number = "30",
    pages = "2450122",
    year = "2024"
}

@article{Haeffner:2008jjg,
    author = "Haeffner, H. and Roos, C. F. and Blatt, R.",
    title = "{Quantum computing with trapped ions}",
    eprint = "0809.4368",
    archivePrefix = "arXiv",
    primaryClass = "quant-ph",
    doi = "10.1016/j.physrep.2008.09.003",
    journal = "Phys. Rept.",
    volume = "469",
    number = "4",
    pages = "155--203",
    year = "2008"
}

@article{Monroe:2019asq,
    author = "Monroe, C. and others",
    title = "{Programmable quantum simulations of spin systems with trapped ions}",
    eprint = "1912.07845",
    archivePrefix = "arXiv",
    primaryClass = "quant-ph",
    doi = "10.1103/RevModPhys.93.025001",
    journal = "Rev. Mod. Phys.",
    volume = "93",
    number = "2",
    pages = "025001",
    year = "2021"
}

@article{Browaeys:2020kzz,
    author = "Browaeys, Antoine and Lahaye, Thierry",
    title = "{Many-body physics with individually controlled Rydberg atoms}",
    eprint = "2002.07413",
    archivePrefix = "arXiv",
    primaryClass = "cond-mat.quant-gas",
    doi = "10.1038/s41567-019-0733-z",
    journal = "Nature Phys.",
    volume = "16",
    number = "2",
    pages = "132--142",
    year = "2020"
}

@article{Saffman:2016kig,
    author = "Saffman, M.",
    title = "{Quantum computing with atomic qubits and Rydberg interactions: progress and challenges}",
    eprint = "1605.05207",
    archivePrefix = "arXiv",
    primaryClass = "quant-ph",
    doi = "10.1088/0953-4075/49/20/202001",
    journal = "J. Phys. B",
    volume = "49",
    number = "20",
    pages = "202001",
    year = "2016"
}

@article{Kitaev:1997wr,
    author = "Kitaev, A. Yu.",
    title = "{Fault tolerant quantum computation by anyons}",
    eprint = "quant-ph/9707021",
    archivePrefix = "arXiv",
    doi = "10.1016/S0003-4916(02)00018-0",
    journal = "Annals Phys.",
    volume = "303",
    pages = "2--30",
    year = "2003"
}

@ARTICLE{Nayak:2008,
       author = {{Nayak}, Chetan and {Simon}, Steven H. and {Stern}, Ady and {Freedman}, Michael and {Das Sarma}, Sankar},
        title = "{Non-Abelian anyons and topological quantum computation}",
      journal = {Reviews of Modern Physics},
         year = 2008,
        month = jul,
       volume = {80},
       number = {3},
        pages = {1083-1159},
          doi = {10.1103/RevModPhys.80.1083},
archivePrefix = {arXiv},
       eprint = {0707.1889},
 primaryClass = {cond-mat.str-el},
       adsurl = {https://ui.adsabs.harvard.edu/abs/2008RvMP...80.1083N}
}

@article{Haeffner:2005bpb,
    author = "Haeffner, H. and others",
    title = "{Scalable multiparticle entanglement of trapped ions}",
    eprint = "quant-ph/0603217",
    archivePrefix = "arXiv",
    doi = "10.1038/nature04279",
    journal = "Nature",
    volume = "438",
    number = "7068",
    pages = "643--646",
    year = "2005"
}

@article{Mirrahimi:2013sgk,
    author = "Mirrahimi, Mazyar and Leghtas, Zaki and Albert, Victor V. and Touzard, Steven and Schoelkopf, Robert J. and Jiang, Liang and Devoret, Michel H.",
    title = "{Dynamically protected cat-qubits: a new paradigm for universal quantum computation}",
    eprint = "1312.2017",
    archivePrefix = "arXiv",
    primaryClass = "quant-ph",
    doi = "10.1088/1367-2630/16/4/045014",
    journal = "New J. Phys.",
    volume = "16",
    number = "4",
    pages = "045014",
    year = "2014"
}

@article{Muhonen:2014gsv,
    author = "Muhonen, Juha T. and others",
    title = "{Storing quantum information for 30 seconds in a nanoelectronic device}",
    eprint = "1402.7140",
    archivePrefix = "arXiv",
    primaryClass = "cond-mat.mes-hall",
    doi = "10.1038/nnano.2014.211",
    journal = "Nature Nanotech.",
    volume = "9",
    pages = "986--991",
    year = "2014"
}

@article{Veldhorst:2014tsi,
    author = "Veldhorst, M. and others",
    title = "{An addressable quantum dot qubit with fault-tolerant control-fidelity}",
    doi = "10.1038/nnano.2014.216",
    journal = "Nature Nanotech.",
    volume = "9",
    pages = "981--985",
    year = "2014"
}

@misc{javadiabhari2024quantumcomputingqiskit,
      title={Quantum computing with Qiskit}, 
      author={Ali Javadi-Abhari and Matthew Treinish and Kevin Krsulich and Christopher J. Wood and Jake Lishman and Julien Gacon and Simon Martiel and Paul D. Nation and Lev S. Bishop and Andrew W. Cross and Blake R. Johnson and Jay M. Gambetta},
      year={2024},
      eprint={2405.08810},
      archivePrefix={arXiv},
      primaryClass={quant-ph},
      url={https://arxiv.org/abs/2405.08810}, 
}

@article{PhysRevLett.82.2417,
  title = {Dynamical Decoupling of Open Quantum Systems},
  author = {Viola, Lorenza and Knill, Emanuel and Lloyd, Seth},
  journal = {Phys. Rev. Lett.},
  volume = {82},
  issue = {12},
  pages = {2417--2421},
  numpages = {0},
  year = {1999},
  month = {Mar},
  publisher = {American Physical Society},
  doi = {10.1103/PhysRevLett.82.2417},
  url = {https://link.aps.org/doi/10.1103/PhysRevLett.82.2417}
}

@article{Temme:2016vkz,
    author = "Temme, Kristan and Bravyi, Sergey and Gambetta, Jay M.",
    title = "{Error Mitigation for Short-Depth Quantum Circuits}",
    eprint = "1612.02058",
    archivePrefix = "arXiv",
    primaryClass = "quant-ph",
    doi = "10.1103/physrevlett.119.180509",
    journal = "Phys. Rev. Lett.",
    volume = "119",
    number = "18",
    pages = "180509",
    year = "2017"
}

@INPROCEEDINGS{9259940,
  author={Giurgica-Tiron, Tudor and Hindy, Yousef and LaRose, Ryan and Mari, Andrea and Zeng, William J.},
  booktitle={2020 IEEE International Conference on Quantum Computing and Engineering (QCE)}, 
  title={Digital zero noise extrapolation for quantum error mitigation}, 
  year={2020},
  volume={},
  number={},
  pages={306-316},
  doi={10.1109/QCE49297.2020.00045}}

@article{Wallman:2015uzh,
    author = "Wallman, Joel J. and Emerson, Joseph",
    title = "{Noise tailoring for scalable quantum computation via randomized compiling}",
    eprint = "1512.01098",
    archivePrefix = "arXiv",
    primaryClass = "quant-ph",
    doi = "10.1103/PhysRevA.94.052325",
    journal = "Phys. Rev. A",
    volume = "94",
    number = "5",
    pages = "052325",
    year = "2016"
}

@article{PhysRevA.105.032620,
  title = {Model-free readout-error mitigation for quantum expectation values},
  author = {van den Berg, Ewout and Minev, Zlatko K. and Temme, Kristan},
  journal = {Phys. Rev. A},
  volume = {105},
  issue = {3},
  pages = {032620},
  numpages = {8},
  year = {2022},
  month = {Mar},
  publisher = {American Physical Society},
  doi = {10.1103/PhysRevA.105.032620},
  url = {https://link.aps.org/doi/10.1103/PhysRevA.105.032620}
}

@article{Bleau:inprep,
  title = "Quantum Entanglement in Collective Neutrino Oscillations
    -- the Role of Quantum Computing",
  author = "Bleau, Katarina and Ilic, Nikolina and Kopp, Joachim and Ramahan, Ushak
    and Yu, Xin Yue",
  year = "2026",
  note = "in preparation",
  journal = "",
}

@inproceedings{Stavenger:2022wzz,
    author = "Stavenger, Timothy J. and Crane, Eleanor and Smith, Kevin C. and Kang, Christopher T. and Girvin, Steven M. and Wiebe, Nathan",
    title = "{C2QA - Bosonic Qiskit}",
    booktitle = "{26th IEEE High Performance Extreme Computing}",
    eprint = "2209.11153",
    archivePrefix = "arXiv",
    primaryClass = "quant-ph",
    doi = "10.1109/HPEC55821.2022.9926318",
    month = "9",
    year = "2022"
}

@article{Araz:2024dcy,
    author = "Araz, Jack Y. and Grau, Matt and Montgomery, Jake and Ringer, Felix",
    title = "{Hybrid quantum simulations with qubits and qumodes on trapped-ion platforms}",
    eprint = "2410.07346",
    archivePrefix = "arXiv",
    primaryClass = "quant-ph",
    reportNumber = "JLAB-THY-24-4200",
    doi = "10.1103/kbv4-jj51",
    journal = "Phys. Rev. A",
    volume = "112",
    number = "1",
    pages = "012620",
    year = "2025"
}

@article{Heeres:2015cnj,
    author = "Heeres, Reinier W. and Vlastakis, Brian and Holland, Eric and Krastanov, Stefan and Albert, Victor V. and Frunzio, Luigi and Jiang, Liang and Schoelkopf, Robert J.",
    title = "{Cavity State Manipulation Using Photon-Number Selective Phase Gates}",
    eprint = "1503.01496",
    archivePrefix = "arXiv",
    primaryClass = "quant-ph",
    doi = "10.1103/physrevlett.115.137002",
    journal = "Phys. Rev. Lett.",
    volume = "115",
    number = "13",
    pages = "137002",
    year = "2015"
}
%=============================================================================

\end{document}